\documentclass{elsart}
\usepackage{epsfig}
\usepackage[figuresright]{rotating}
\begin{document}

\newcommand{\bc}{\begin{center}}
\newcommand{\ec}{\end{center}}
\newcommand\cc{{\rm CC}}
\newcommand\nc{{\rm NC}}
\newcommand\nutau{\ensuremath{\nu_\tau\,}}
\newcommand\anutau{\ensuremath{\bar\nu_\tau\,}}
\newcommand\numu{\ensuremath{\nu_\mu\,}}
\newcommand\anumu{\ensuremath{\bar\nu_\mu\,}}
\newcommand\nue{\ensuremath{\nu_e\,}}
\newcommand\anue{\ensuremath{\bar\nu_e\,}}

\newcommand\enunu{\ensuremath{e^- \anue \nu_\tau\,}}
\newcommand\mununu{\ensuremath{\mu^- \anumu \nu_\tau\,}}
\newcommand\taue{\ensuremath{\tau^- \to \enunu\,}}
\newcommand\taumu{\ensuremath{\tau^- \to \mununu\,}}
\newcommand\onepr{\ensuremath{h^- (n\pi^0) \nutau\,}}
\newcommand\tauonepr{\ensuremath{\tau^- \to h^- (n\pi^0) \nutau\,}}
\newcommand\taupi{\ensuremath{\tau^- \to h^- \nutau\,}}
\newcommand\taurho{\ensuremath{\tau^- \to \rho^- \nutau\,}}
\newcommand\threepi{\ensuremath{h^-h^+h^-\,}}
\newcommand\tauthreepi{\ensuremath{\tau^- \to \threepi \nutau\,}}
\newcommand\tauthreepr{\ensuremath{\tau^- \to \threepi (n\pi^0) \nutau\,}}
\newcommand\threepr{\ensuremath{\threepi (n\pi^0) \nutau\,}}
\newcommand\tauplusthreepi{\ensuremath{\tau^+ \to \pi^+\pi^-\pi^+ \anutau\,}}

\newcommand\numunutau{\ensuremath{\numu \to \nutau\,}}
\newcommand\numunue{\ensuremath{\numu \to \nue\,}}
\newcommand\nuenutau{\ensuremath{\nue \to \nutau\,}}
\newcommand\anumunutau{\ensuremath{\anumu \to \anutau\,}}
\newcommand\epsmu{\ensuremath{\epsilon_\mu\,}}
\newcommand\epstau{\ensuremath{\epsilon_\tau\,}}
\newcommand\Eetot{\ensuremath{E_{\rm e}^{\rm tot}\,}}
\newcommand\evis{\ensuremath{E_{\rm vis}\,}}
\newcommand\qlep{\ensuremath{Q_{\rm Lep}\,}}
\newcommand\pt{\ensuremath{p_T\,}}
\newcommand\lcc{\ensuremath{l_{\rm CC}\;}} 
\newcommand\lmucc{\ensuremath{l^{\mu}_{\rm CC}\;}} 
\newcommand\lecc{\ensuremath{l^{e}_{\rm CC}\;}} 
\newcommand\xvis{\ensuremath{\tau_{\rm V}\,}}
\newcommand\vpxvis{\ensuremath{\mathbf p^{\,\xvis}\,}}
\newcommand\jet{\ensuremath{H}}                    
\newcommand\ybj{\ensuremath{y_{Bj}}               }

\newcommand\mpigamgam{\ensuremath{M_{\pi\gamma\gamma}}} 
\newcommand\mthreepi{\ensuremath{M_{3\pi}\,}}

\newcommand\Nmuobs{\ensuremath{N_\mu^{\rm obs}\,}}
\newcommand\ntaumax{\ensuremath{N_\tau^{\rm max}\,}}
\newcommand\ntaumu{\ensuremath{N_\tau^{\mu \tau }\,}}
\newcommand\ntaue{\ensuremath{N_\tau^{e \tau }\,}}

\newcommand\define{\;\; \equiv \;\;}
\newcommand\mycal{\mathcal}
\newcommand\liksel{\ensuremath{\mycal{L}^{\rm S}\,}}
\newcommand\likv{\ensuremath{\mycal{L}^{\rm V}\,}}
\newcommand\likint{\ensuremath{\mycal{L}^{\rm IN}\,}}
\newcommand\liknc{\ensuremath{\mycal{L}^{\rm NC}\,}}
\newcommand\likcc{\ensuremath{\mycal{L}^{\rm CC}\,}}

\newcommand\ratsel{\ensuremath{\ln\lambda^{\rm S}\,}}
\newcommand\ratv{\ensuremath{\ln\lambda^{\rm V}\,}}
\newcommand\ratint{\ensuremath{\ln\lambda^{\rm IN}\,}}
\newcommand\ratnc{\ensuremath{\ln\lambda^{\rm NC}\,}}
\newcommand\ratcc{\ensuremath{\ln\lambda^{\rm CC}\,}}

\newcommand\qtrat{\ensuremath{R_T\,}}

\newcommand\epsdir{.}

\newcommand{\fix}[1]{{\bf <<< #1 !!! }}

\begin{frontmatter}
\title{\bf Final NOMAD results on \boldmath{\numunutau} and 
\boldmath{\nuenutau} oscillations including a new search for 
\boldmath{$\nu_{\tau}$} appearance using hadronic \boldmath{$\tau$} decays.} 
\collab{NOMAD Collaboration}
\author[Paris]             {P.~Astier}
\author[CERN]              {D.~Autiero}
\author[Saclay]            {A.~Baldisseri}
\author[Padova]            {M.~Baldo-Ceolin}
\author[Paris]             {M.~Banner}
\author[LAPP]              {G.~Bassompierre}
\author[Lausanne]          {K.~Benslama} 
\author[Saclay]            {N.~Besson}
\author[CERN,Lausanne]     {I.~Bird}
\author[Johns Hopkins]     {B.~Blumenfeld}
% ...

\address[LAPP]           {LAPP, Annecy, France}                               
\address[Johns Hopkins]  {Johns Hopkins Univ., Baltimore, MD, USA}            
% ...
\address[CERN]           {CERN, Geneva, Switzerland}                          
\address[Lausanne]       {University of Lausanne, Lausanne, Switzerland}      
% ...
\address[Padova]         {Univ. of Padova and INFN, Padova, Italy}            
\address[Paris]          {LPNHE, Univ. of Paris VI and VII, Paris, France}    
% ...
\address[Saclay]         {DAPNIA, CEA Saclay, France}                         
% ...

\clearpage
\begin{abstract}
Results from the $\nu_{\tau}$ appearance search   
in a neutrino beam using the full NOMAD data sample are reported. 
A new analysis unifies all the hadronic $\tau$ decays, 
significantly improving the overall sensitivity of the 
experiment to oscillations. The ``blind analysis" of 
all topologies yields no evidence for an oscillation signal. 
In the two-family oscillation scenario,  
this sets a 90\% C.L. allowed region in the 
$\sin^22\theta_{\mu\tau} - \Delta m^2$ plane which 
includes $\sin^22\theta_{\mu\tau}\ <\ 3.3\times10^{-4}$ at 
large $\Delta m^2$ and $\Delta m^2 < 0.7$ eV$^2$/$c^4$ at 
$\sin^22\theta_{\mu \tau}=1$. The corresponding contour in the 
$\nuenutau$ oscillation hypothesis results in 
$\sin^22\theta_{e\tau}\ <\ 1.5\times10^{-2}$ at large $\Delta m^2$ 
and $\Delta m^2 < 5.9$ eV$^2$/$c^4$ at $\sin^22\theta_{e \tau}=1$.  
We also derive limits on effective couplings of the $\tau$ lepton 
to \numu or \nue. 
\end{abstract}
\end{frontmatter}


\section{Introduction}
\label{sec:intro}

The NOMAD experiment was designed in 1991 to search for \nutau 
appearance from neutrino oscillations in the CERN wide-band 
neutrino beam produced by the 450 GeV proton synchrotron. 
Given the neutrino energy spectrum and the distance between 
the neutrino source and the detector ($\sim 600$ m on average), 
the experiment is sensitive to differences of mass 
squares $\Delta m^2 > 1$~eV$^2$. 

The experiment was motivated by theoretical arguments 
suggesting that the \nutau may have a mass of 1 eV, or larger, 
and therefore could be the main constituent of the 
dark matter in the universe. 
This suggestion was based on two assumptions: 
\begin{itemize} 
\item the interpretation of the solar neutrino deficit~\cite{solar} 
      in terms of $\nu_{e} \to \nu_{\mu}$ oscillations amplified 
      by matter effects~\cite{matter}, giving 
      $\Delta m^{2} \approx 10^{-5}$ eV$^{2}$; 
\item the so-called ``see-saw" model~\cite{seesaw} which predicts 
      that neutrino masses are proportional to the square 
      of the mass of the charged lepton, or of the charge 2/3 
      quark of the same family. 

\end{itemize} 
From these two assumptions one expects a \numu mass of 
$\sim 3\times 10^{-3}$ eV and a \nutau mass of $\sim 1$ eV, or higher. 
Furthermore, in analogy with quark mixing, neutrino mixing angles 
were expected to be small. 

It is within this theoretical scenario that the NOMAD 
experiment~\cite{NOMADNIM} started a direct search 
for \numunutau oscillations together with another 
experiment (CHORUS) which used the same neutrino 
beam but searched for \nutau appearance with a  
complementary technique~\cite{chorus}.  


The detection of an oscillation signal in NOMAD 
relies on the identification of \nutau 
charged-current (CC) interactions using kinematic criteria.
The spatial resolution of the detector does not resolve 
the $\tau$ decay vertex from the \nutau CC interaction. 
The identification of \nutau CC events 
is thus achieved by exploiting all the kinematic constraints 
which can be constructed from a precise measurement of all 
visible final-state particles.  
This requires a detector with good energy and momentum 
resolution and sophisticated analysis schemes. 
The NOMAD experiment is the first neutrino oscillation 
search to use this technique~\cite{Albright}. 

From 1995 to 1998 the experiment has collected 1~040~000 
events with an identified muon, corresponding to about 1~350~000 
$\numu$ CC interactions, given the combined trigger, vertex 
identification, and muon detection efficiency~of~77\%.  

A recent paper~\cite{PLB1} described a search for \numunutau and 
\nuenutau oscillations in the full NOMAD data sample. The analysis 
was based on both deep inelastic (DIS) interactions and low-multiplicity 
(LM) events for all the accessible $\tau$ decay channels. 
No evidence for oscillations was observed.  

In this paper we report an improved $\nu_{\tau}$ appearance 
search in the hadronic $\tau$ decay channels.  
We then combine this search with the analyses of the 
$\nu_{\tau}e\bar{\nu}_{e}$ DIS channel and of the LM topologies 
described in Ref.~\cite{PLB1} and we present the overall NOMAD results. 
The new analysis unifies all the hadronic $\tau$ decay topologies 
($\nu_{\tau}h$, $\nu_{\tau}\rho$ and $\nu_{\tau}3h$) 
in a single selection scheme. 
A refined implementation of kinematics, exploiting all 
the available degrees of freedom, together with new algorithms 
for the rejection of CC backgrounds leads to an improvement of 
the combined NOMAD sensitivity to oscillations by almost a 
factor of two with respect to Ref.~\cite{PLB1}. 
The systematic uncertainties are also substantially reduced.

\section{NOMAD detector}
\label{sec:det}

\begin{figure}[tb]
\bc
  \epsfig{file=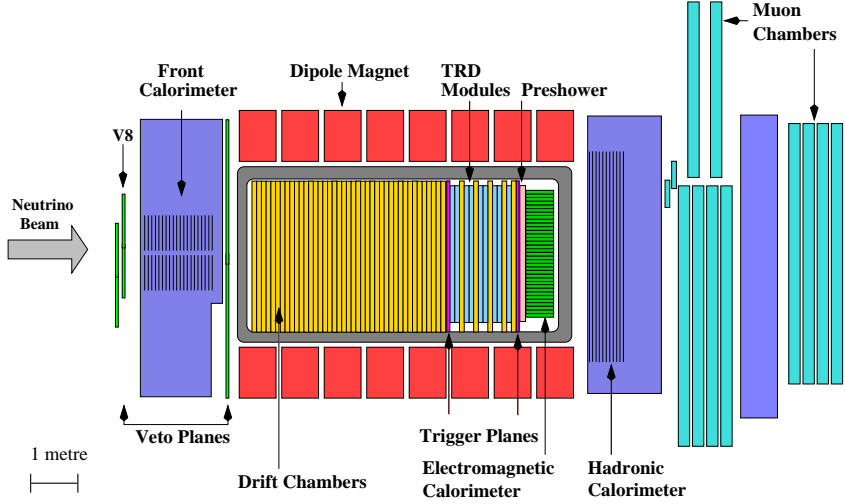,angle=-90,width=0.80\textwidth}
\ec
\caption{Side view of the NOMAD detector. 
}
\label{fig:det}
\end{figure}

The NOMAD detector (Figure~\ref{fig:det}) is described 
in detail in Ref.~\cite{NOMADNIM}. Inside
a 0.4 T magnetic field there is an active target consisting of
drift chambers (DC)~\cite{DC} with a fiducial mass of about 2.7 tons 
and a low average density (0.1 g/cm$^3$). The target, 405 cm 
long and corresponding to about one radiation length,
is followed by a transition radiation detector (TRD)~\cite{TRD} 
for electron identification, a preshower detector (PRS), and    
a high resolution lead-glass electromagnetic
calorimeter (ECAL)~\cite{ECAL}. A hadron calorimeter (HCAL) and 
two stations of drift chambers for muon detection are located 
just after the magnet coil.
The detector is designed to identify leptons and to measure 
muons, pions, electrons and photons with comparable resolutions.
Momenta are measured in the DC with a resolution:  
$$
\frac{\sigma_p}{p}\simeq \frac{0.05}{\sqrt{L[m]}}\oplus 
\frac{0.008 \times p[GeV/c]}{\sqrt{L[m]^5}}
$$
where L is the track length and $p$ is the momentum. 
The energy of electromagnetic showers, $E$, is measured 
in the ECAL with a resolution: 
$$
\frac{\sigma_E}{E}=0.01 \oplus \frac{0.032}{\sqrt{E[GeV]}}.
$$

The neutrino interaction trigger \cite{trigger}
consists of a coincidence between signals from two
planes of counters located after the active target, in the absence of
a signal from a large area veto system in front of the
NOMAD detector.

\section{Neutrino beam and event samples}
\label{sec:beam}

From recent beam computations~\cite{nutaubeam}, 
the relative composition of CC events in NOMAD is 
estimated to be \numu CC\,:\,\anumu CC\,:\,\nue CC\,:\,\anue CC=
1.00\,: \,0.0227\,: \,0.0154: \,0.0016, with average
neutrino energies of 45.4, 40.8, 57.5, and 51.5 GeV, respectively.\
The prompt $\nutau$ component is negligible \cite{nutaucont}.
Neutrinos are produced at an average distance of 625 m from the detector.
In addition to the \numu CC events, the data contain about 34~000 $\anumu$ CC, 
21~000 $\nue$ CC, 2300 $\anue$ CC interactions, and  
485~000 neutral current (NC) interactions. 
 
Neutrino interactions are simulated using a modified versions of
LEPTO 6.1 \cite{Lepto} and JETSET 7.4 \cite{Jetset} with $Q^2$
and $W^2$ cutoff parameters removed, and with $\tau$ mass and
polarization effects included. We use the nucleon Fermi motion
distribution of Ref.~\cite{Bodek}, truncated at 1 GeV/$c$. 
To define the parton content of the nucleon for the cross-section 
calculation we use the GRV-HO parametrization~\cite{GRV} of 
the parton density functions, available in PDFLIB~\cite{PDFLIB}.
A full detector simulation based on GEANT~\cite{Geant} is performed.
Further corrections to these samples are applied using 
the data themselves, as described in Section~\ref{sec:ds}. 
The size of the simulated samples exceeds the data sample by a factor
of about 3 for \numu CC interactions, 10 for NC
and \anumu CC interactions and 100 for \nue and \anue CC 
interactions. In addition, more than 500~000 $\nutau$ CC
events have been generated for each $\tau^-$ decay channel.

\section{Analysis principles}
\label{sec:principles}

\begin{figure}[tb]
\bc
  \epsfig{file=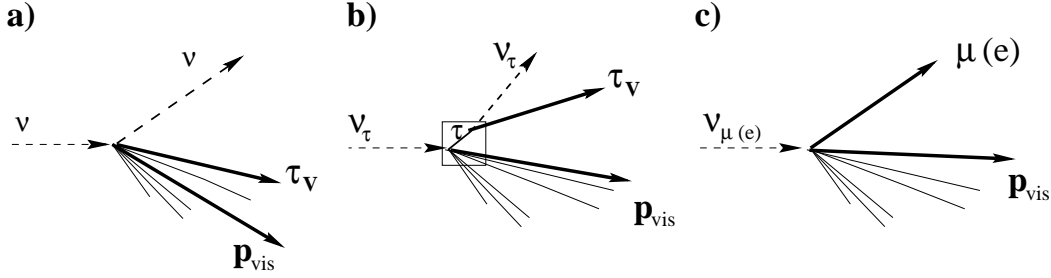,width=1.00\textwidth}
\ec
\caption{
Signal and background topologies in NOMAD: a) NC background; 
b) \nutau CC signal with subsequent $\tau$ decay; c) 
$\nu_{\mu}(\nu_{e})$ CC background. The square indicates 
the reconstructed ``primary" vertex for \nutau CC interactions.  
The effect of the \xvis selection on $\nu_{\mu}(\nu_{e})$ CC 
topologies is discussed in Sections~\ref{sec:veto} and~\ref{sec:intst}. 
}
\label{fig:bkgnds}
\end{figure}

From the kinematical point of view, \nutau CC events 
in NOMAD are fully characterized by the (undetected) decay 
of the primary $\tau$. The presence of visible secondary 
$\tau$ decay products, \xvis, distinguishes them from NC 
interactions, whereas the emission of one(two) neutrino(s) in 
hadronic(leptonic) $\tau$ decays provides discrimination 
against \numu (\nue) CC interactions (Figure~\ref{fig:bkgnds}). 
Consequently, in \nutau CC events the transverse component 
of the total visible momentum and the variables 
describing the visible decay products 
have different absolute values and different correlations 
with the remaining hadronic system, $H$, than in \numu 
(\nue) CC and NC interactions.
The optimal separation between signal and background 
is achieved when all the degrees of freedom of the 
event kinematics (and their correlations) are exploited.  

A rejection power against backgrounds of $\mathcal{O}(10^5)$ 
is required from the kinematic analysis in order to 
match the data sample size (Section~\ref{sec:beam}).  
In addition, the potential \nutau signal allowed by 
limits from previous experiments \cite{E531,CDHS} is at least 
by a factor of 0.0025 times smaller than the main \numu CC component.  
Therefore, the \nutau appearance search in NOMAD 
is a search for rare events within large  
background samples. This imposes severe constraints 
on the analysis techniques. In order to obtain 
reliable background estimates we have developed 
methods to correct Monte Carlo (MC) predictions 
with experimental data and we have defined appropriate 
control samples to check our predictions.

This paper describes a new search for $\nutau$ CC interactions 
in the hadronic $\tau$ decay channels $\onepr$ and $\threepr$, 
for a total branching ratio of 64.7\%~\cite{PDG}. The analysis 
focuses on DIS events, defined by 
a cut on the total hadronic momentum recoiling against the 
visible $\tau$ decay product(s), $p^{H}>1.5$ GeV/c.  

Neutrino interactions in the active target are selected
by requiring the presence of at least one charged track in addition 
to the potential $\tau$ decay products, originating from a 
common vertex in the detector fiducial volume.
Quality cuts are then applied to ensure that 
the selected events are properly reconstructed.  
This is obtained by imposing constraints on the statistical 
significance of the main kinematic variables 
(see Appendix~\ref{sec:variables} and Section~\ref{sec:kine}): 
$\sigma(p_T^{\,m})/p_T^{\,m}$ (transverse plane), $\sigma(\qlep)/\qlep$, 
$\sigma(p^{\xvis})/p^{\xvis}$ and $\sigma_{\rm max}(p)/p^{\xvis}$ 
($[\mathbf{p}^{\xvis},\mathbf{p}^{H}]$ plane).  
In addition, loose requirements based on approximate charge balance 
at the primary vertex are also applied. Overall, these 
quality cuts typically remove 10 to 15\% of the events.  

The separation of \nutau CC interactions from backgrounds 
is described in detail in Section~\ref{sec:scheme} and can be 
summarized in the following steps (Figure~\ref{fig:kinevar}): 
\begin{itemize} 
\item the visible $\tau$ decay products \xvis are 
      identified on the basis of their topology (Section~\ref{sec:sele}); 

\item constraints are applied to the structure of the  
      hadronic system $H$ recoiling against \xvis (Section~\ref{sec:jet});  

\item specific algorithms are used to identify and veto primary  
      leptons originating from \numu (\nue) CC interactions 
      (Section~\ref{sec:veto});  

\item constraints are imposed on the internal structure 
      of the selected \xvis can\-di\-da\-te (Section~\ref{sec:intst}); 

\item the final background rejection is achieved by exploiting 
      all the available information from the global 
      event topology (Section~\ref{sec:kine}).  
\end{itemize}  
Section~\ref{sec:bkgrel} is devoted to the background 
estimate, with a description of the data corrections  
(Section~\ref{sec:corr}) and of the control samples 
used to check background predictions (Section~\ref{sec:control}).  
Systematic errors are discussed in Section~\ref{sec:syserr}. 
The final results from the analysis of the hadronic DIS channels 
are then combined in Section~\ref{sec:results} with the 
remaining topologies from Ref.~\cite{PLB1}.  

In the following we describe the main analysis principles, 
which have been extensively discussed in Ref.~\cite{PLB1}\cite{PLB2}.

\subsection{Statistical analysis of data}
\label{sec:like} 

The kinematic variables used in the present analysis are 
defined in Ap\-pen\-dix~\ref{sec:variables}. 
In order to exploit their correlations, these variables 
are combined into likelihood functions. 
These functions are partial two-, three- or four-dimensional 
(2D, 3D, 4D) probability density functions ({\em pdf}), 
denoted in the following by square brackets (e.g. $[a,b]$ denotes 
the 2D correlation between the variables $a$ and $b$).  
The final likelihood function, $\mycal{L}$, is then obtained from a  
combination of partial pdf's which includes correlations 
(denoted again by square brackets) among them.  
This analysis technique provides the best sensitivity 
to oscillations. As is common practice, the logarithm of the final 
likelihood ratio between test hypotheses, $\ln \lambda$, is used.  

A fit to $\ln\lambda$ with separate signal and background 
components can increase the statistical power of the search. 
However, since few events are expected in the signal region,  
we combine regions of similar signal to background 
ratio into suitably chosen bins of variable 
size (Section~\ref{sec:kine}). 
These bins are then treated as statistically 
independent (Section~\ref{sec:limits}).

\subsection{Data simulator}
\label{sec:ds}

In order to reliably compute signal and background efficiencies, 
the MC results are corrected using the data 
themselves (Section~\ref{sec:corr}).
We perform this correction by using a sample of $\numu$ CC events from 
the data, removing the {\em identified} muon, and replacing it 
with a MC-generated lepton $\ell$ of appropriate momentum vector, 
where $\ell$ can be a $\nu$, an $e^-$ or a $\tau^-$ followed by its decay. 
In addition, as explained in Section~\ref{sec:corr}, events 
with identified muons or electrons are directly used to correct 
CC background predictions in the hadronic $\tau$ decays.  
All these samples are referred to as the Data Simulator (DS). 

The same procedure is then applied to reconstructed MC $\numu$ CC 
events and these event samples are referred to as the 
Monte Carlo Simulator (MCS). In order to reduce systematic 
uncertainties, signal and background efficiencies $\epsilon$ 
are then obtained from the relation:  
$$ 
\epsilon = \epsilon_{\rm MC} \times \epsilon_{\rm DS} / \epsilon_{\rm MCS}
$$ 
which implies that efficiencies for lepton reconstruction are 
obtained from the MC ($\epsilon_{\rm MC}$), while the 
effect of the hadronic jet differences between 
data and simulations is taken into account through the 
ratio $\epsilon_{\rm DS} / \epsilon_{\rm MCS}$.  
It has been checked that efficiencies obtained by 
this method are indeed stable with respect to 
variations of the MC input models (nuclear effects, 
fragmentation, Fermi motion, detector resolution functions) 
within the quoted uncertainty. 

The errors on background predictions given in all the 
following tables reflect the statistical uncertainties 
from MC, MCS and DS. Systematic 
uncertainties are discussed in Section~\ref{sec:syserr}.

\subsection{Avoiding biases} 
\label{sec:biases} 

A procedure referred to as ``blind analysis" is used. 
According to this procedure, data events inside the signal 
region (the ``blind box") cannot 
be analyzed until all the selection criteria are defined and 
the robustness of the background predictions is demonstrated 
with appropriate control samples (Section~\ref{sec:control}). 
The selection criteria are chosen by optimizing 
the sensitivity to oscillations (Section~\ref{sec:kine}). 
This is defined as the average upper limit on the oscillation probability  
that would be obtained, in the absence of a signal, by an ensemble 
of experiments with the same expected background \cite{STAT}.  

The sensitivity to oscillations also provides the final criterion 
for any choice between different analyses of the same $\tau$ 
decay topology or for the replacement of a previous analysis 
by a newer one. The choice is made before looking at data 
events falling in the signal region. 

The MC samples used to define the selection 
criteria are different from those used to evaluate the 
background and signal efficiency, which are therefore 
fully unbiased.

\section{Selection scheme} 
\label{sec:scheme} 

The new selection of hadronic $\tau$ decays unifies 
four specific decay topologies in a single general scheme: 
\begin{itemize} 
\item[$0\gamma$:] {$\tau \rightarrow \nu_{\tau}\pi$ decays. The \xvis 
                  candidate is built from a single charged track.} 
\item[$1\gamma$:] {$\tau \rightarrow \nu_{\tau}\rho \rightarrow  
                  \nu_{\tau}\pi \pi^{0}$ decays 
                  where the $\pi^{0}$ is reconstructed as a single 
                  ECAL cluster either because the two photons 
                  overlap in ECAL or because one of them is not 
                  reconstructed in ECAL.  
                  The \xvis candidate is built from a charged 
                  track and from a single ECAL neutral cluster.}  
\item[$2\gamma$:] {$\tau \rightarrow \nu_{\tau}\rho \rightarrow 
                  \nu_{\tau}\pi \pi^{0}$ decays  
                  where the $\pi^{0}$ is reconstructed from two 
                  separate ECAL clusters. The \xvis candidate is 
                  built from a charged track and from two ECAL 
                  neutral clusters.} 
\item[$3h$:]      {$\tau \rightarrow \nu_{\tau}a_{1} \rightarrow 
                  \nu_{\tau}\pi\rho^{0} \rightarrow 
                  \nu_{\tau}\pi\pi\pi$ decays. The \xvis 
                  candidate is built from three charged tracks.} 
\end{itemize}  
Each topology is independently analyzed  
with the {\em same selection criteria}. 
Events selected by more than one topology define a further 
sample. Eventually, we combine the different topologies 
into a single search. This statistical treatment provides 
the best overall sensitivity to oscillations. 
Photon conversions in the DC volume are not used  
for $\pi^0$ reconstruction in this analysis. However, events where  
only one of the $\gamma$'s from $\pi^0$ decay converts into an 
$e^+e^-$ pair are included in the $1\gamma$ topology.  
The signal efficiencies quoted in the following refer to the 
$\nu_{\tau}h(n\pi^0)$ (Br 49.5\%) or $\nu_{\tau} 3h(n\pi^0)$ 
(Br 15.2\%)~\cite{PDG} inclusive DIS samples. 
 
The selection scheme is intended to exploit, at each step, all 
the available topological information (degrees of freedom) 
through the use of appropriate probability density functions 
based on correlations among kinematic variables. 
As explained in Appendix~\ref{sec:variables}, an 
event in NOMAD can be fully described by five degrees of 
freedom (Figure~\ref{fig:kinevar}): 
three in the transverse plane $(x,y)$
and two along the beam direction $(z)$.

\subsection{Identification of $\tau$ decay product(s)} 
\label{sec:sele} 

In hadronic $\tau$ decays the selection of the visible 
decay product(s) relies on topological 
constraints. This implies that, for a {\em given event}, 
more than one choice is possible and 
therefore a discriminating criterion is needed.    
Since the total visible momentum of the event 
($\evis$ and $p_{T}^{m}$) is fixed  
in the problem, only three degrees of freedom related to the  
general event structure are available. In addition, when applicable, 
the internal structures of $\xvis$ and $H$ provide further constraints.  

\begin{sidewaysfigure}[p]
\bc
  \epsfig{file=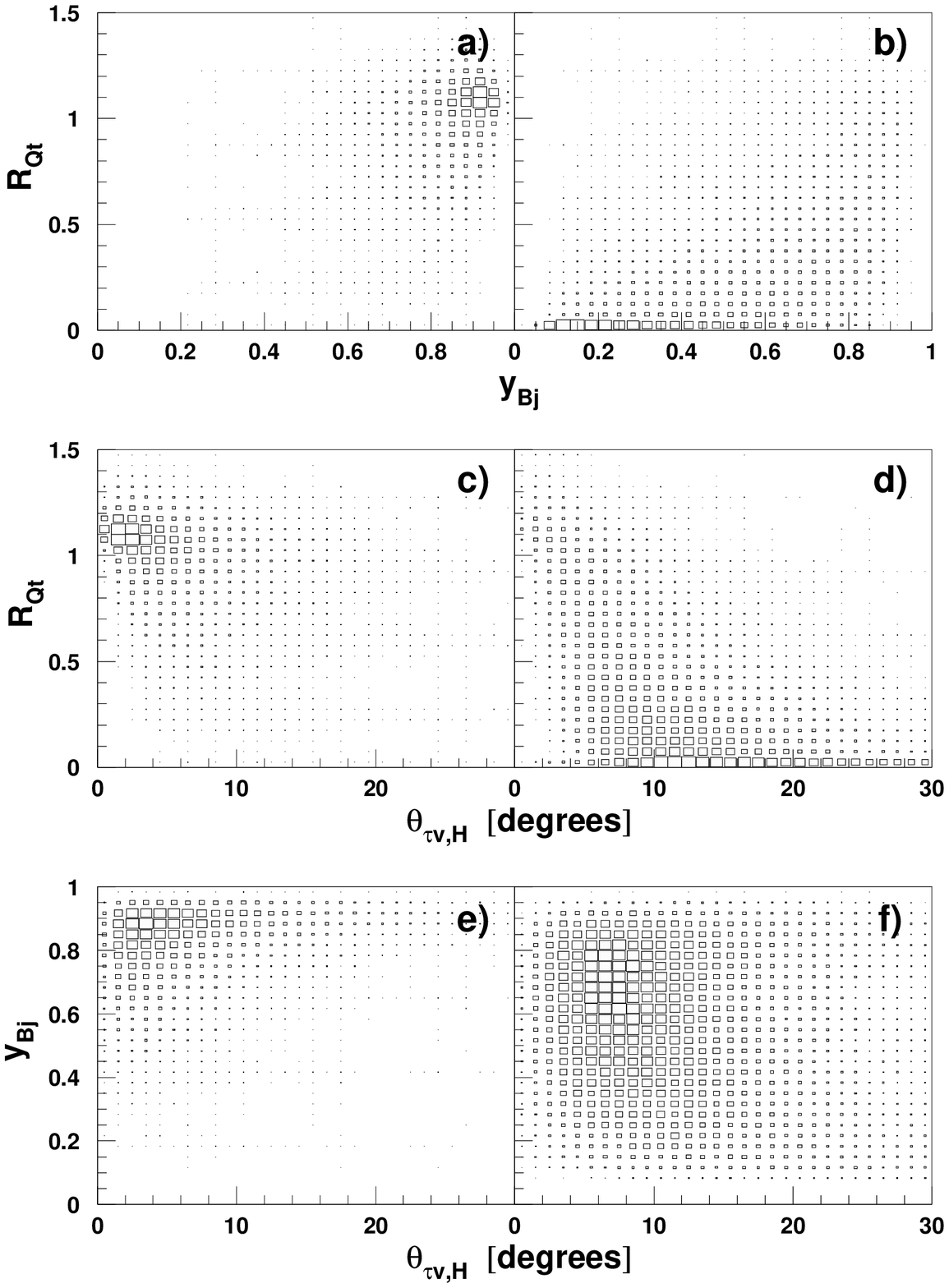,width=0.50\textwidth}\epsfig{file=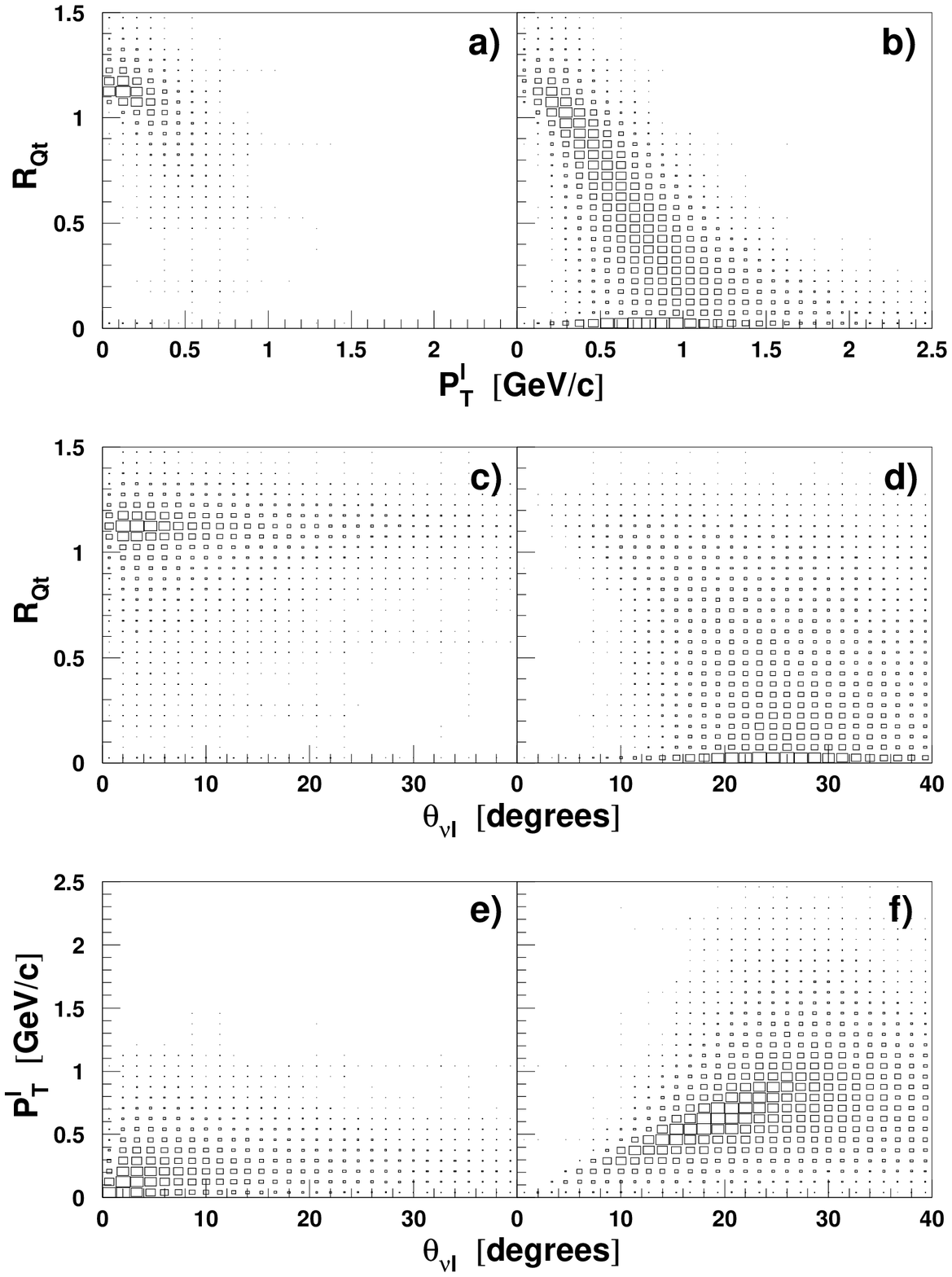,width=0.50\textwidth}
\ec
\caption{
Left plots: correlations between kinematic variables, not related 
to the internal \xvis structure, used to construct \ratsel for 
wrong (a,c,e) and correct (b,d,f) combinations of $\tau$ 
decay products in $\nu_{\tau}\pi$ decays (Section~\ref{sec:sele}). 
Right plots: correlations between kinematic variables used to 
construct \ratv for wrong tracks (a,c,e) and for the true 
leading muon (b,d,f) in $\nu_{\mu}$ CC events passing the first 
level veto (Section~\ref{sec:veto}). 
}
\label{fig:tag_vars}
\end{sidewaysfigure}

A single likelihood function, $\liksel$, is used to select 
among all the combinations the most likely $\xvis$ originating 
from a $\tau$ decay: 
$$ 
\liksel \; \equiv \; [\; (\likint), R_{Q_{T}}, y_{Bj}, 
\theta_{\tau_{\rm V} H} \;]   
$$ 
The 2D correlations between the three variables $R_{Q_{T}}$, $y_{Bj}$ 
and $\theta_{\xvis H}$ (Appendix~\ref{sec:variables}) 
are shown in Figure~\ref{fig:tag_vars} (left plots). 
The function $\likint$ describes the internal $\xvis$ structure for the 
$1\gamma$, $2\gamma$ and 3h topologies (obviously, it cannot be 
defined for the $0\gamma$ topology): 
$$ 
\likint \;\; \equiv \;\; 
\left\{ 
\begin{array}{rl} 
 [\;M_{\rho},\theta_{\pi^{-}\pi^0},E_{\pi^0}/E_{\rm vis}\;] & \;\;\;\;\; 1\gamma  \\ 

[[\;M_{\pi^0},\theta_{\gamma\gamma},E_{\gamma}^{\rm max}/E_{\rm vis}\;], M_{\rho},\theta_{\pi^{-}\pi^0},E_{\pi^0}/E_{\rm vis}\;] & \;\;\;\;\; 2\gamma  \\ 

[[\;\overline{M}_{\rho^{0}}\;,\overline{\theta}_{\pi^+\pi^-},E_{\pi^+} \;\;\;\;\;], M_{a_{1}},\theta_{\pi^-\pi^-},\overline{E}_{\rho^{0}} \;\;\;\;\;\;]  & \;\;\;\;\; 3h \\ 
\end{array} \right. 
$$ 
where, for the $3h$ topology, the average of the two indistinguishable 
$\pi^+\pi^-$ combinations ($\overline{M}_{\rho^{0}}, 
\overline{\theta}_{\pi^+\pi^-},\overline{E}_{\rho^{0}}$) is used. 
This function is based on the decay kinematics of the $\rho$ 
($1\gamma$ and $2\gamma$) and $a_{1}$ (3h) resonances in 
the laboratory frame.  
In the definition of \likint the invariant masses 
are considered as free pa\-ra\-me\-ters.  

The isolation variable $R_{Q_{T}}$ is sensitive to the internal 
structure of the hadronic system and thus incorporates  
additional information with respect to the degrees of freedom 
related to the global event topology. 
The use of this variable for the \xvis identification results, 
overall, in four degrees of freedom in addition to the 
\xvis internal structure. However, only three of them are 
physically relevant for the \xvis identification and are 
therefore included in \liksel. This procedure (and the 
choice of the variables entering \liksel) also minimizes  
potential biases induced on background events, in which  
a fake (not coming from $\tau$ decay) \xvis is indeed 
constructed by the \xvis identification algorithm itself.  

A likelihood ratio, $\ratsel$, is built as the ratio of the $\liksel$ 
function for correct and wrong combinations of $\tau$ decay products 
in $\nu_{\tau}$ CC events. In each event, the system of particles with 
maximum $\ratsel$ among all possible combinations is tagged as $\xvis$.  
No cut is applied on $\ratsel$, which is constructed from 
{\em signal events} only and, therefore, is not the optimal 
discriminant against backgrounds (Section~\ref{sec:intst}).  
After the $\xvis$ identification, a minimum $\gamma$ energy of 
200 MeV (100 MeV) is required for the $1\gamma(2\gamma)$ topology  
in order to remove fake photons. 
The algorithm correctly identifies the visible $\tau$ decay product(s) 
in 91\%, 80\%, 80\% and 73\% of all the $0\gamma$, $1\gamma$, $2\gamma$ 
and $3h$ $\nu_{\tau}$ CC events respectively. 
These values set an approximate upper limit to the final 
selection efficiency, since events with a mis-identified \xvis 
do not have the kinematics expected from 
\nutau CC events (Section~\ref{sec:kine}).

\subsection{Structure of the $H$ system} 
\label{sec:jet} 

For background events the $\xvis$ candidate is mostly selected 
inside the hadronic jet. In NC interactions the whole visible 
event is indeed a hadronic jet (Fi\-gu\-re~\ref{fig:bkgnds}a). 
On the other hand, in CC interactions where the leading lepton 
is not identified and the \xvis particle(s) are embedded in the jet, 
the remaining hadronic system $H$ has an incorrect 
topology (Figure~\ref{fig:mucckine}b). As a consequence,   
a constraint on the structure of the hadronic system 
significantly increases the background rejection. 

In order to effectively reject both NC and CC topologies, the 
jet structure must include information from $\xvis$.  
This is achieved by defining the variable: 
$$ 
S_{H} \;\; \equiv \;\; \frac{\langle Q_{T}^{2} \rangle_{H}}
{\Delta r_{\tau_{\rm V} h_{i}}^{2/3}}  
$$ 
which takes into account the transverse size of the 
hadronic system, $\langle Q_{T}^{2} \rangle_{H}$, and the 
opening of the minimum invariant cone between $\xvis$ and any 
other charged track, $\Delta r_{\xvis h_{i}}$. 
The requirement $S_{H}<0.20(0.16)$ GeV$^2/c^2$ 
in the one (three) prong search selects di-jet topologies.  
Due to the higher average multiplicity, the $3h$ topology is more 
sensitive to variables related to the internal structure of $\xvis$ 
(Section~\ref{sec:intst}) and $H$, thus requiring a tighter constraint. 

An additional check of the $H$ structure is obtained by 
removing $\xvis$ and by computing the maximum transverse 
momentum \qlep among all 
charged tracks within the hadronic system. As explained in 
Section~\ref{sec:elveto}, events originating from  
$\nu_{e}(\bar{\nu}_{e})$ CC interactions are indeed characterized by 
large values of \qlep for the {\em leading lepton}.  
The condition $Q_{\rm Lep}^{\rm max} < 5.0$~GeV/c further 
suppresses topologies where the unidentified leading 
electron (positron) is erroneously included in the hadronic 
system (Figure~\ref{fig:mucckine}b).

\subsection{Lepton veto} 
\label{sec:veto} 

\begin{figure}[tb]
\bc
  \epsfig{file=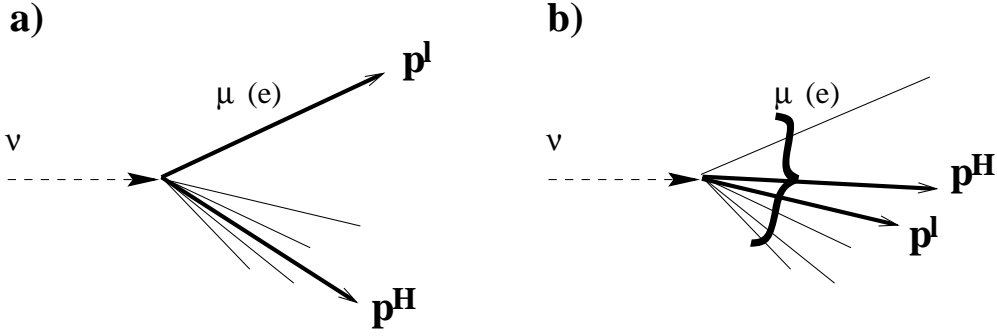,width=0.95\textwidth}
\ec
\caption{
Possible topologies for an unidentified $\nu_{\mu}(\nu_{e})$ CC 
interaction after
the selection of a leading charged track in the analysis:
a) the primary lepton is chosen as the leading  
particle; b) the primary lepton is included in the 
hadronic system $H$. Similar considerations can be applied 
to the \xvis selection ($l$ is replaced by \xvis). 
However, for the $1\gamma$, $2\gamma$ and $3h$ samples
the choice of the primary lepton as one of the \xvis 
particles can produce additional topologies, 
as explained in Section~\ref{sec:intst} and shown 
in Figure~\ref{fig:ccint}.  
}
\label{fig:mucckine}
\end{figure}

Due to the presence of a highly isolated track (leading lepton), 
{\em unidentified} CC interactions can fake hadronic $\tau$ decays.  
As discussed above, 
the selection of a single leading charged track defines two 
different topologies in CC background events, depending  
on whether the leading lepton is the chosen track  
(Figure~\ref{fig:mucckine}a) or is included in the 
remaining hadronic system $H$ (Figure~\ref{fig:mucckine}b).  
In the latter case the transverse plane kinematics 
is significantly distorted by the selection of 
a random leading particle, thus reducing their 
effectiveness. Specific algorithms are then developed 
in order to tag the leading lepton. 
The selection can be divided in two steps: 
\begin{itemize} 
\item{{\em First level veto} (FLV). \\  
      Lepton identification criteria are used 
      to reject events containing clear primary leptons of 
      any charge. 
} 
\item{{\em Second level veto} (SLV). \\  
      A kinematic criterion based on event topology is 
      used to tag a {\em single} lepton candidate, \lmucc or \lecc, 
      among all negatively charged primary tracks. Subsequently, 
      additional lepton identification criteria are imposed 
      to the selected track.  
} 
\end{itemize}  
This method aims at avoiding tight lepton identification 
criteria. In particular, it is also possible to analyze 
events outside the geometrical acceptance of the relevant 
subdetectors (Figure~\ref{fig:det}). 
The analyses described in Ref.~\cite{PLB1}  
were instead based on tight identification conditions 
and rejected all events containing high $p_{T}$ 
tracks escaping the detector. This resulted in a 
significant efficiency loss. The kinematic tagging 
plays a crucial role in the veto algorithm because events 
in which the leading lepton is correctly tagged are  
efficiently rejected by kinematics (Section~\ref{sec:kine}).

\begin{sidewaystable}[p]
\caption{
Number of observed events and expected background  
as a function of cuts for all the hadronic DIS decay modes. The 
corresponding signal efficiency (\%) is listed in the first 
column ($\nutau$ CC). The effect of the selection 
on the $\tau^+$ control sample is also shown. 
The numbers are fully corrected with the data simulator and the last column  
shows the ratio between data and predicted background. 
The efficiencies refer to the inclusive $h(n\pi^{0})$ and 
$3h(n\pi^{0})$ samples and do not include branching ratios. 
Overlap events are included in each topology.  
} 
\vspace*{0.2cm}
\bc
\begin{tiny}
\begin{tabular}{lcrcrrcrrcrrcrrcrrccc}
\hline 
Sample & & \multicolumn{1}{c}{$\nutau$ CC} && \multicolumn{2}{c}{NC} && 
\multicolumn{2}{c}{$\nu_{\mu}$+$\bar{\nu}_{\mu}$ CC} && \multicolumn{2}{c}{$\nue$+$\bar{\nu}_{e}$ CC} && \multicolumn{2}{c}{Tot Bkgnd} &&
\multicolumn{2}{c}{Data} && \multicolumn{2}{c}{Data/Bkgnd} \\ \cline{5-6} \cline{8-9} \cline{11-12} \cline{14-15} \cline{17-18} \cline{20-21} 
Charge & & \multicolumn{1}{c}{--} && \multicolumn{1}{c}{--} & \multicolumn{1}{c}{+} && \multicolumn{1}{c}{--} & \multicolumn{1}{c}{+} && \multicolumn{1}{c}{--} & \multicolumn{1}{c}{+} && \multicolumn{1}{c}{--} & \multicolumn{1}{c}{+} && \multicolumn{1}{c}{--} & \multicolumn{1}{c}{+} && \multicolumn{1}{c}{--} & \multicolumn{1}{c}{+} \\ \hline
 $\xvis$ candidate& $0\gamma$ & 43.7 && 80739 & 88663  && 27532  & 30852  && 1612  & 1426  && 109883 & 120941 && 114012 & 125013 && 1.04 & 1.03  \\
  &  $1\gamma$ & 43.1 && 95869  & 103264  && 25531  & 29094  && 1670 & 1591  && 123070  & 133949  && 126128  & 137516  && 1.02  & 1.03   \\
  &  $2\gamma$ & 33.7 && 79934  & 80613  && 21247  & 25742  && 1358  & 1494  && 102539  & 107849  && 105399  &  109285 && 1.03  & 1.01  \\
  &  $3h$ & 44.4  && 105524 & 116237     && 37874  & 41406  && 1612  &  2452 && 145060 & 160095  && 148420 & 165094  && 1.02 & 1.03 \\\cline{1-21}
 $H$ structure & $0\gamma$ & 36.3 && 68383  & 78517   && 23987  & 23420  && 1305  & 940  && 93675  & 102877   && 96020   & 105938  && 1.02 & 1.03  \\
  & $1\gamma$ & 36.5 && 81682  & 90164   && 20347   & 22476  && 1233   & 1002   && 103262  & 113642   && 104609  & 115531  && 1.01 & 1.02  \\
  & $2\gamma$ & 27.7 && 66257  & 68570   && 16533   & 19187   && 976  & 917  && 83766  & 88674  && 85102   &  89317 && 1.02 & 1.01  \\
  & $3h$ & 32.8 && 88472  & 98737 && 28794 &31171  &&1179  &1576  &&118445  &131484  && 116106  & 134205  && 0.98 & 1.02 \\\cline{1-21}
 Lepton veto & $0\gamma$ & 23.7 && 39319  & 40077  && 7422  & 16299  && 165  & 389  && 46906  & 56765  && 47673  & 54986  && 1.02 & 0.97  \\
   &  $1\gamma$ & 17.5 && 32128   & 32260  && 3930   & 5598   && 164  & 289   && 36222  & 38147   && 36818  & 39066 && 1.02 & 1.02  \\
   &  $2\gamma$ & 12.6 && 23569  & 22800  && 2847  & 4345   && 120   & 244  && 26536  & 27389   && 27158  & 27913  && 1.02 & 1.02  \\
  & $3h$ & 13.2 && 19701  & 25712 && 2196 & 2566 && 182 & 381 && 22079 & 28659  && 21929 & 27889  && 0.99 & 0.97 \\\cline{1-21}
 $\ratint$ cut &  $0\gamma$ &--&&--&--&&--&--&&--&--&&--&--&&--&--&&--&--\\
  &  $1\gamma$ & 9.5 && 8568  & 7888  && 718  & 1499  && 69  & 125   && 9355 & 9512  && 9212  & 9408  && 0.98 & 0.99 \\
  &  $2\gamma$ & 7.1 && 8702   & 8048  && 843  & 1515  && 59  & 113  && 9604  & 9676 && 9840  & 9951  && 1.02 & 1.03 \\
  & $3h$ & 4.9 && 2411 & 2366 &&147&219 &&28 &45&&2586&2630 && 2666  & 2684  && 1.03 & 1.02 \\\cline{1-21}
 $\ratcc$ cut &  $0\gamma$ & 9.7 && 12020   & 11464  && 717  & 684  && 55  & 88  && 12792 & 12236 && 13471  & 12101  && 1.05 & 0.99 \\
  &  $1\gamma$ & 8.6 && 7816  & 7459 && 565  & 1127  && 62  & 107  && 8443 & 8693 && 8345  & 8696  && 0.99 & 1.00 \\
  &  $2\gamma$ & 6.2 &&  7864  & 7509  && 616  & 1137 && 51  & 95  && 8531  & 8741  && 8789  & 9097  && 1.03 & 1.04 \\
  & $3h$ & 4.2  && 2079  & 2025 &&119&173&&20&43&&2219&2241&& 2253  & 2243  && 1.02 & 1.00 \\\cline{1-21}
 $\ratnc$ cut &  $0\gamma$ & 1.3 && 6.7$\pm$2.1  & 7.3$\pm$2.5  && 0.4$\pm$0.4 & 2.0$\pm$0.6 && 0.7$\pm$0.1  & 4.0$\pm$0.3  && 7.8$\pm$2.2 & 13.3$\pm$2.6  && 10  & 11  && 1.28$\pm$0.54 & 0.83$\pm$0.30 \\
  &  $1\gamma$ & 1.4 && 6.7$\pm$2.1   & 8.9$\pm$3.1  && 0.5$\pm$0.4  & 3.5$\pm$0.9  && 0.9$\pm$0.1  & 4.6$\pm$0.4  && 8.1$\pm$2.2  & 17.0$\pm$3.2  &&  8 & 17  && 0.99$\pm$0.44 & 1.00$\pm$0.31 \\
  &  $2\gamma$ & 0.8 && 3.4$\pm$1.3 & 7.9$\pm$1.7  && 1.1$\pm$0.6  & 3.4$\pm$0.9  && 0.9$\pm$0.2  & 3.3$\pm$0.4  && 5.4$\pm$1.5  & 14.6$\pm$2.0  && 6  & 16  && 1.11$\pm$0.55 & 1.10$\pm$0.31 \\
  & $3h$ & 1.3  && 2.8$\pm$1.4  & 5.3$\pm$1.2 && 1.5$\pm$0.6 &
  2.6$\pm$0.9 && 0.6$\pm$0.2 &2.0$\pm$0.3&& 4.9$\pm$1.5 &9.9$\pm$1.6 && 3  & 10  && 0.61$\pm$0.40 & 1.01$\pm$0.36 \\
\hline 
\end{tabular} 
\end{tiny}
\label{tab:hadrons} 
\ec
\end{sidewaystable}

\subsubsection{Muon tagging and veto} 
\label{sec:muveto} 

The FLV rejects events containing a $\mu^{-}(\mu^{+})$ positively 
identified by the reconstruction algorithms or by the presence of 
residual hits and track segments associated to a charged 
track in both stations of the muon chambers. 

The muon tagging in the SLV is based on the likelihood function: 
$$
\likv \define [\; R_{Q_{T}}, p_{T}^{l}, \theta_{\nu l} \;] 
$$ 
which is similar to $\liksel$ but takes into account the topology 
of $\nu_{\mu}$ CC interactions with unidentified muons. 
These events are characterized, in general, by large 
values of $y_{Bj}$, and the effect of variables 
relating the hadronic system to the muon track is weaker. 
For this reason, $\likv$ only includes variables directly 
describing properties of the track being considered. 
The 2D correlations between variables used to construct 
\likv are shown in Figure~\ref{fig:tag_vars} (right plots). 

A likelihood ratio, $\ratv$, is built from $\nu_{\mu}$ CC events 
{\em surviving the FLV}, as the ratio of $\likv$ for  
the true muon and for other tracks. In the three prong search $\lmucc$  
is defined as the track with maximum $\ratv$ 
among all the negatively charged primary tracks.  
In the one prong search the tagging efficiency can be further increased 
by considering the bias induced by the $\xvis$ selection, which 
returns a single charged track. In this case the \xvis 
charged track is also tagged as $\lmucc$ if,   
for this track, $\ratv > 0$, otherwise $\lmucc$ is chosen 
as the track with maximum $\ratv$ (the most likely muon). 

The algorithm correctly identifies the true muon in 
94\%, 89\%, 89\% and 86\% of $\nu_{\mu}$ CC surviving the 
FLV for the $0\gamma$, $1\gamma$, $2\gamma$ and $3h$ topologies.  
In addition, the muons not identified by the algorithm 
have very low energy and  
are thus generally included in the hadronic 
system $H$ (Figure~\ref{fig:mucckine}b). This results in 
a NC-like configuration which is efficiently rejected by 
kinematics against NC interactions (Section~\ref{sec:kine}).  

Only events in which $\lmucc$ is compatible with being a minimum 
ionizing particle in both ECAL and HCAL are rejected at 
this stage. The suppression of the remaining \numu CC 
background is achieved by exploiting the complete 
event kinematics under the assumption that 
{\em \lmucc is the leading particle} (see Appendix~\ref{sec:variables}),  
as explained in Section~\ref{sec:kine}.

\subsubsection{Electron veto} 
\label{sec:elveto} 

Although \nue(\anue) CC events represent only a tiny 
fraction of the full neutrino interactions (Section~\ref{sec:beam}), 
\nue(\anue) can be potentially more dangerous than \numu CC events. 
This is due to the lower identification efficiency of 
electrons with respect to muons (limited angular 
acceptance of TRD, efficiency loss for $\pi$ rejection, 
bremsstrahlung emission, etc.) and to the fact that 
this effect is not restricted to specific topologies.  
In addition, \nue have a harder energy spectrum, 
because they originate mainly from $K$ decays, 
and therefore the primary lepton is well isolated 
from the hadronic jet. However, in the hadronic $\tau$ decays 
the presence of $\gamma$'s from $\pi^{0}$ (sometimes converting 
in the DC volume) can produce electron-like signals. 
For this reason, the analysis does 
not use stringent constraints against 
$\nu_{e}(\bar{\nu}_{e})$ CC interactions, which would 
result in a significant loss of efficiency. 

The FLV is designed to reject two different classes of events. 
First, events containing a primary track which is {\em positively 
identified} as $e^{-}(e^{+})$ by both TRD and loose PRS requirements 
are rejected if, for this track, $p_{T}^{l}/p_{T}^{m}>1.0$ and 
$p_{T}^{l}>1.0$ GeV/c. These are mainly high $p_{T}$ 
$e^{-}(e^{+})$ in events with small missing transverse momentum. 
A second background source originates from $\nu_{e}(\bar{\nu}_{e})$ 
CC interactions where {\em no particle identification} is 
available because the primary  
electron (positron) emitted most of its energy as a hard bremsstrahlung 
$\gamma$ and missed the relevant subdetectors. Therefore, 
for each primary track (of either charge) not reaching the TRD, we  
search for potential bremsstrahlung secondaries  
with direction tangential to the track. 
The presence of such additional $\gamma$'s or charged particles 
is then used as a kind of ``electron tag" 
and the event is rejected if, for the fully reconstructed 
lepton (sum of primary track and bremsstrahlung secondaries),  
$p_{T}^{l}/p_{T}^{m}>1.5$ and $p_{T}^{l}>1.0$ GeV/c.

Since, contrary to the muon case, the electron identification 
inefficiencies are not restricted to specific topologies, 
the SLV tagging criterion is simply based on $\ratsel$.  
A $\lecc$ candidate is then defined as the \xvis charged track 
in one-prong events and as the negative \xvis  
track with the larger momentum uncertainty in three-prong events. 
Events from $\nu_{e}$ CC interactions 
in which the primary electron is chosen as $\tau$ daughter 
candidate are indeed dangerous because they are 
highly isolated and cannot be suppressed by studying the jet structure.  
This happens in 85\%, 62\%, 57\% and 51\% of all \nue CC events 
for the $0\gamma$, $1\gamma$, $2\gamma$ and $3h$ topologies. 
These events are rejected if the $\lecc$ track fulfills loose 
electron identification criteria based on TRD and on 
the combined PRS-ECAL information.  

As described in the following, the background in the most 
sensitive region of the analysis (Table~\ref{tab:lowbkg}) 
consists almost entirely of $\nu_{e}(\bar{\nu}_{e})$ CC interactions.   
This is partially explained by the kinematic approach 
used (Section~\ref{sec:kine}).

\begin{figure}[tb]
\bc
  \epsfig{file=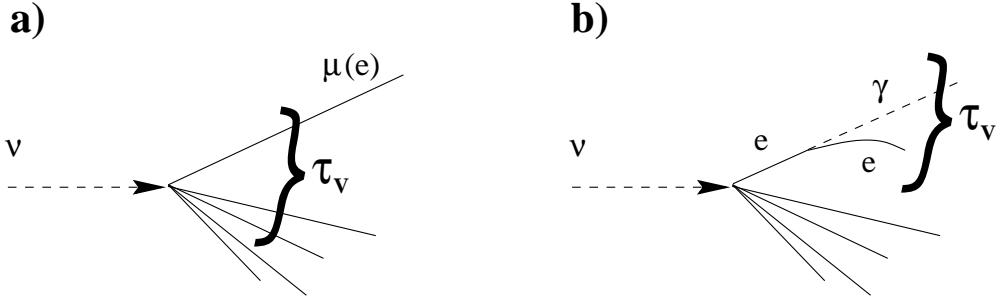,width=0.95\textwidth}
\ec
\caption{
Internal \xvis topologies ($1\gamma$, $2\gamma$ and $3h$) 
for CC backgrounds in which the leading 
lepton is part of the \xvis candidate: a) $\nu_{\mu}(\nu_{e})$ CC 
interaction with part of \xvis embedded in the hadronic jet; 
b) $\nu_{e}$ CC interaction with bremsstrahlung emission.
}
\label{fig:ccint}
\end{figure}

\begin{sidewaysfigure}[p]
\bc
  \epsfig{file=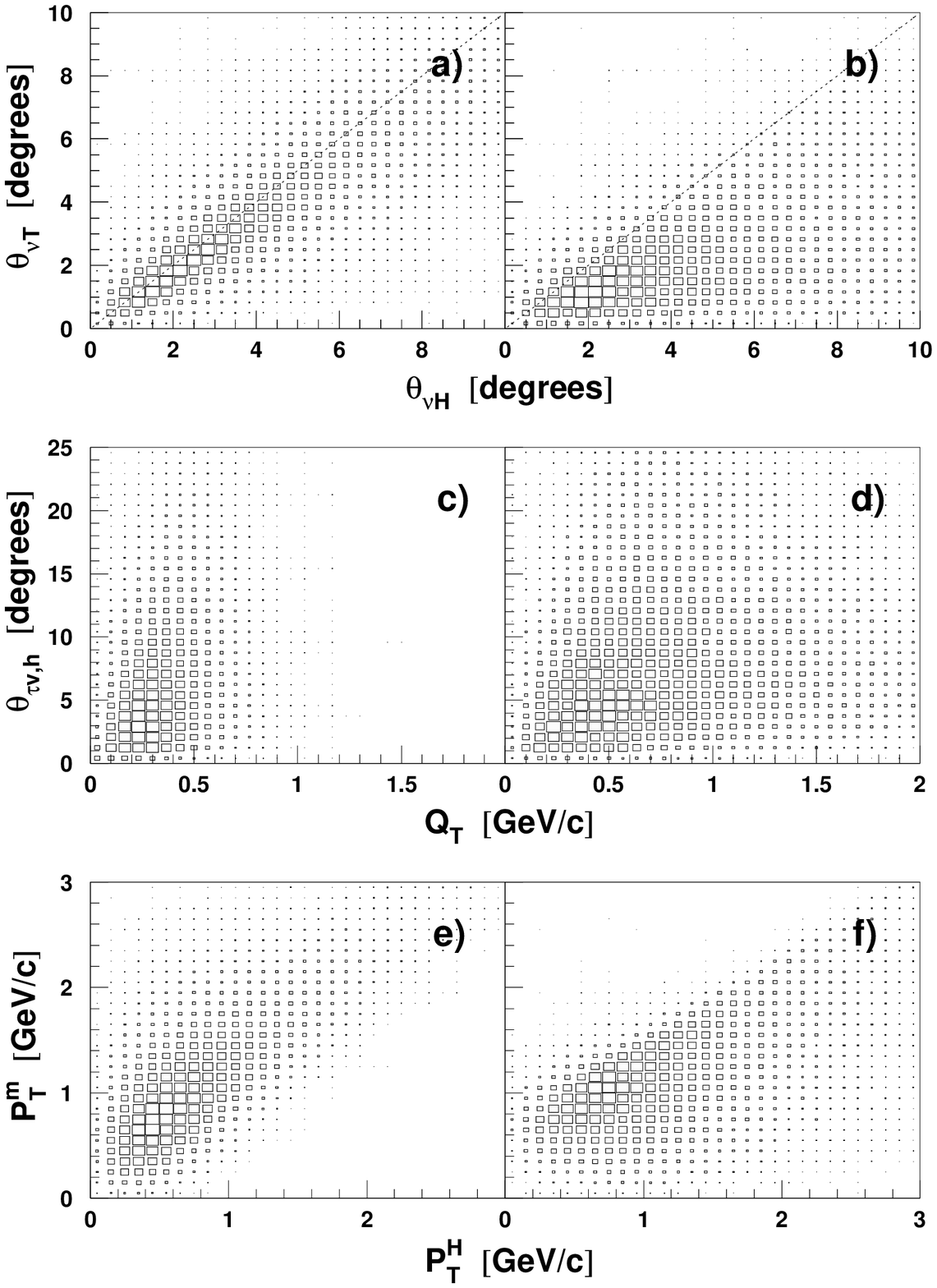,width=0.50\textwidth}\epsfig{file=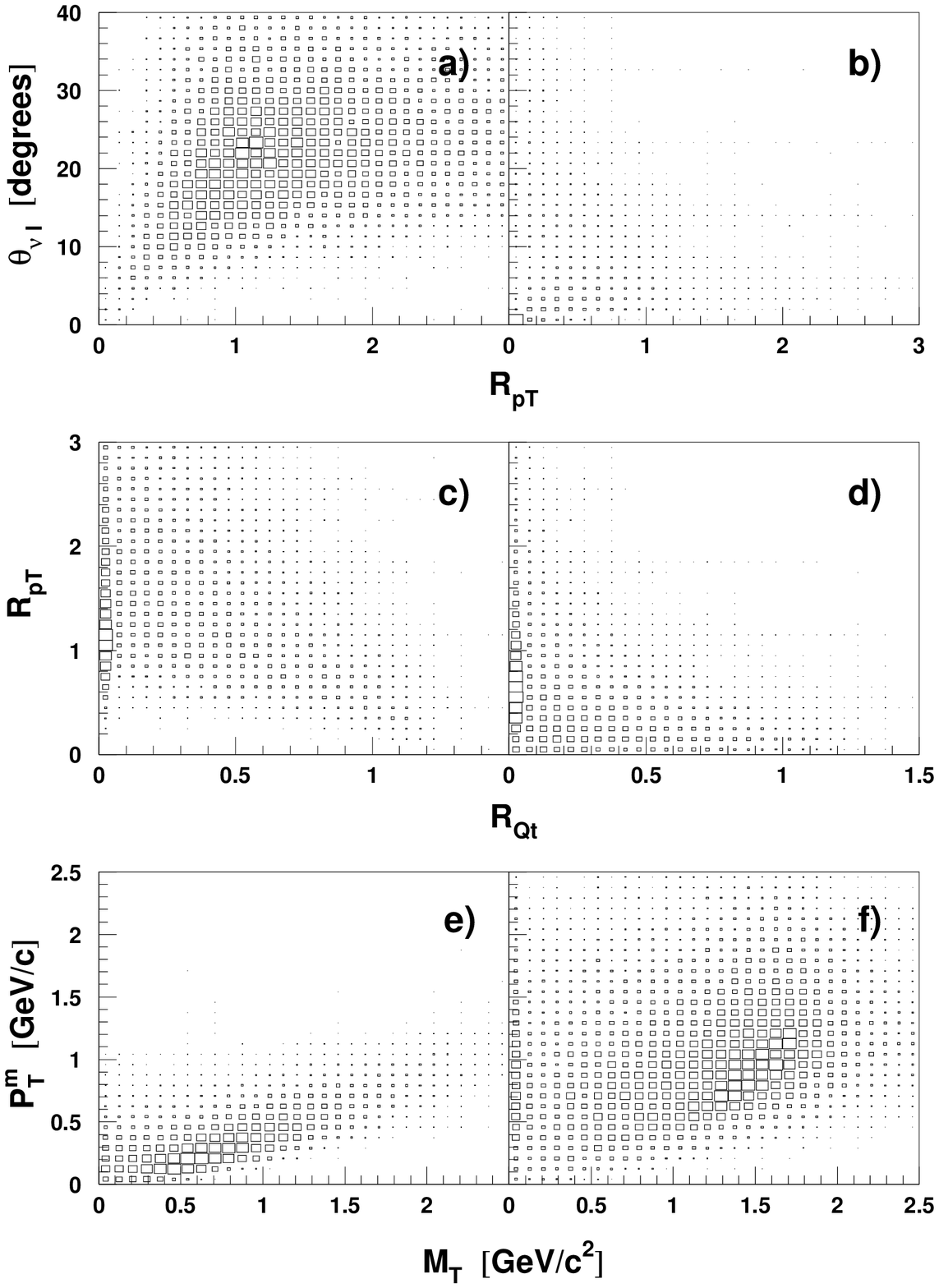,width=0.50\textwidth}
\ec
\caption{
Left plots: correlations between kinematic variables 
used to construct $\ratnc$ for MC events of type $\nu_{\mu}$ NC (a,c,e) 
and $\tau \rightarrow \nu_{\tau}\pi$ decays with the selection 
of the right $\pi$ as leading particle (b,d,f). Right plots: correlations 
between kinematic variables for MC events of type $\nu_{\mu}$ CC 
with unidentified $\mu$ (a,c,e) and $\tau \rightarrow \nu_{\tau}\pi$ decays 
with the chosen \lmucc as leading particle (b,d,f).  
}
\label{fig:tau_vars}
\end{sidewaysfigure}

\subsection{Structure of the $\xvis$ candidate} 
\label{sec:intst} 

For the $1\gamma$, $2\gamma$ and $3h$ topologies the internal 
$\xvis$ structure provides further discrimination against backgrounds.  
The corresponding information, incorporated into the $\likint$ function,  
has been already used for the $\xvis$ identification 
(Section~\ref{sec:sele}), through the likelihood ratio between 
correct and random combinations {\em in signal events}. 

Once the $\xvis$ selection is performed, background events can 
still have a $\xvis$ internal structure inconsistent with $\tau$ 
decay. In particular, the CC sample contains two specific 
topologies. The first one includes events where $\xvis$ consists 
of the leading $\mu(e)$ and of additional particle(s) from 
the hadronic jet (Figure~\ref{fig:ccint}a).  
A second possibility arises when the leading electron from 
a $\nu_{e}$ CC interaction undergoes hard bremsstrahlung   
radiation and $\xvis$ consists of different pieces 
of the original electron (Figure~\ref{fig:ccint}b). 
Both cases can be suppressed by 
a constraint on the $\xvis$ internal structure, which makes 
use of the presence of $\rho$ or $a_{1}$ resonances in $\tau$ decays.  
However, the $\xvis$ internal structure is expected to be 
less effective against NC background because these events 
may contain genuine resonances inside the hadronic jet.  

A likelihood ratio, $\ratint$, is then built as the ratio of 
the $\likint$ function between the true combination in 
signal events and the selected $\xvis$ in a weighted mixture of  
$\nu_{\mu}$ and $\nu_{e}$ CC interactions. A loose cut at 
$\ratint >0.5,0.0,0.0$ is applied for the $1\gamma$, $2\gamma$ 
and $3h$ topologies respectively. 
The higher charged multiplicity of the $3h$ topology 
partially correlates the $\xvis$ internal structure to the jet 
structure (Section~\ref{sec:jet}) for backgrounds and therefore 
a further binning is used along $\ratint$ 
in this case (Section~\ref{sec:kine}). Since no kinematic 
constraint on any variable related to the \xvis structure 
has been previously imposed, the \ratint cut results in a 
significant event reduction for all samples (Table~\ref{tab:hadrons}).  
However, this is due mainly to the rejection of 
events which {\em do not contain} \xvis candidates compatible 
with being a genuine $\rho$ or $a_{1}$ resonance.

\begin{sidewaysfigure}[p]
\bc
  \epsfig{file=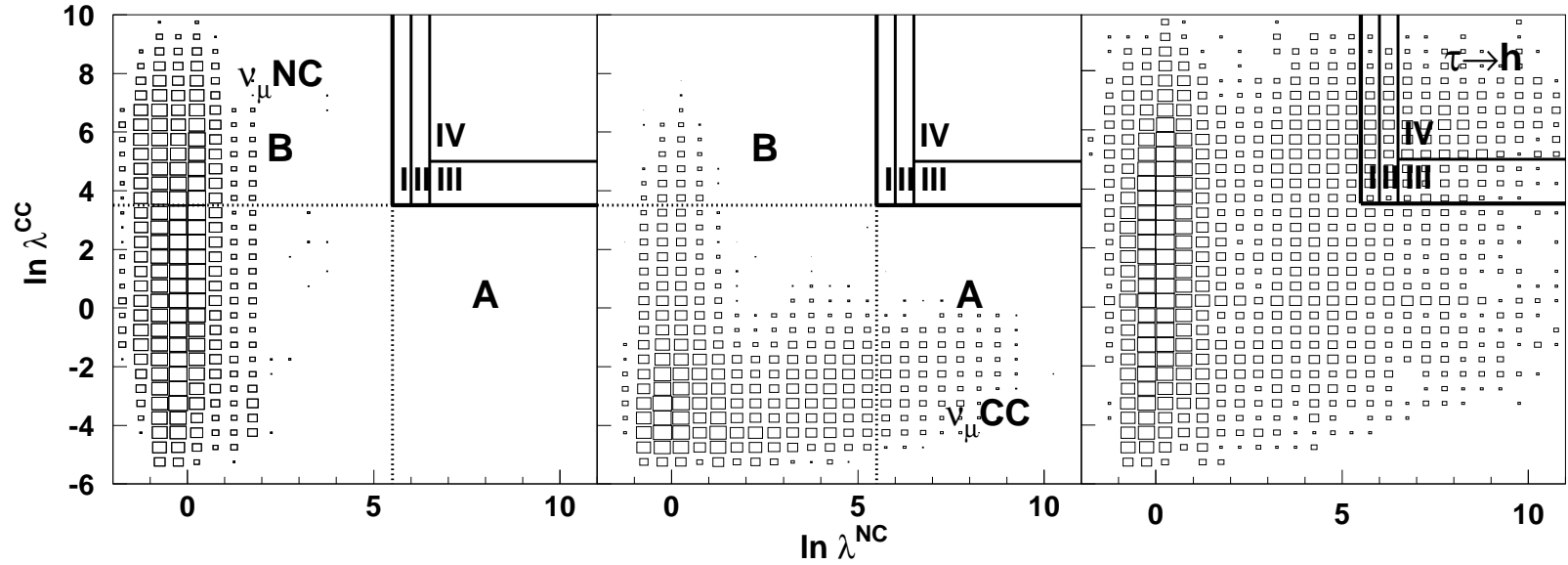,width=0.90\textwidth}
\ec
\caption{ 
Scatter plot of $\ratcc$ vs $\ratnc$ for MC backgrounds from \numu 
NC (left plot) and \numu CC interactions passing the first level veto 
(middle plot) and simulated $\tau \rightarrow \nu_{\tau}\pi$ ($0\gamma$) 
decays (right plot). The local bin densities are shown in logarithmic scale.  
The large box at the upper right corner, divided into four bins, 
defines the signal region, whereas the dotted lines show   
the control regions used to estimate the systematic 
uncertainties on CC (region A) and NC (region B) events. 
}
\label{fig:like2D}
\end{sidewaysfigure}

\subsection{Global kinematics} 
\label{sec:kine} 

Background events can be divided into two categories with 
opposite kinematic configurations. In NC interactions 
{\em the $\xvis$ candidate} is embedded in the hadron 
jet (Figure~\ref{fig:bkgnds}a), and a large missing 
transverse momentum associated with the 
escaping neutrino is almost opposite to the 
direction of the hadronic system. 
On the other hand, {\em the leading lepton} in CC 
interactions is typically well-isolated and balances the 
momentum of the remaining hadronic system 
in the transverse plane (Figure~\ref{fig:bkgnds}c).  
The signal from hadronic $\tau$ decays has intermediate 
properties between these two extremes; 
the $\tau$ decay neutrino introduces a modest amount 
of missing transverse momentum and the non-colinearity 
of $\tau$ and \xvis can reduce the isolation of 
\xvis (Figure~\ref{fig:bkgnds}b).

\begin{sidewaysfigure}[p]
\bc
  \epsfig{file=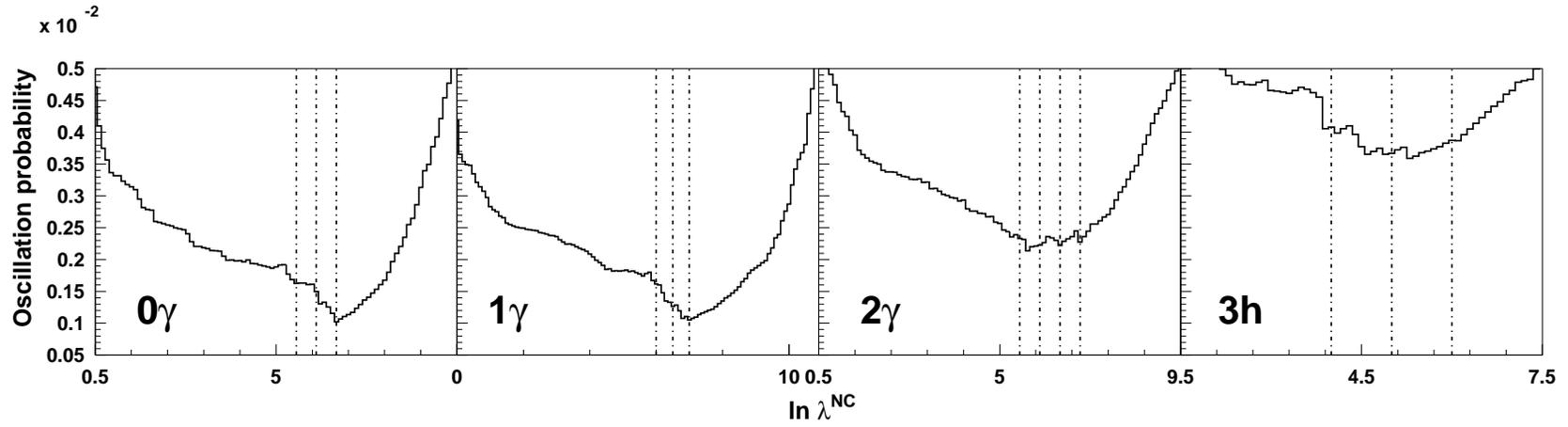,width=1.00\textwidth}
\ec
\caption{
Definition of the signal region: sensitivity as a function of
$\ratnc$ for the $0\gamma$, $1\gamma$, $2\gamma$ and $3h$
(from left to right) topologies. The vertical lines correspond to
the binning used in the analysis. The blind region starts at
$\ratnc > 5.5,6.0,5.5,4.0$ for the $0\gamma$, $1\gamma$,
$2\gamma$ and $3h$ topologies respectively (left vertical
lines in each plot).
A cut $\ratcc > 3.5,0.0,0.0,0.0$ is applied.
}
\label{fig:sensiall}
\end{sidewaysfigure}

In order to optimize separately the rejection of each of the two 
opposite background sources, we implement an event classification 
based on the use of two distinct likelihood functions. 
As discussed in the following, this procedure provides the  
best description of the kinematic information  
and, at the same time, the possibility to distinguish further  
between NC and CC backgrounds. The first likelihood function  
is designed to separate the signal from NC interactions:  
$$
\liknc \define 
[[[\; \theta_{\nu T}, \theta_{\nu H} \;], \theta_{\xvis h_{i}}, Q_{T} \;],
p_{T}^{m}, p_{T}^{H} \;]
$$
where the minimum opening angle $\theta_{\xvis h_{i}}$ is  
sensitive to the internal structure of the hadronic system.  
Figure~\ref{fig:tau_vars} (left plots) illustrates the discriminating 
power of some correlations between the kinematic variables used 
to construct $\liknc$. For each event, a likelihood ratio $\ratnc$ 
is computed as the ratio of the $\liknc$ functions constructed 
from the true $\xvis$ combination in signal events and from the 
selected $\xvis$ combination in NC events respectively.  
Since $\liknc$ is built by selecting {\em \xvis as the leading 
particle}, for the CC background a cut on $\ratnc$  
suppresses the topology where the unidentified lepton 
has little energy and is part of the hadronic system 
$H$ (NC-like configuration in Figure~\ref{fig:like2D}). 
As explained in Section~\ref{sec:muveto}, this reduces 
the mistagging probability of muons to a negligible level 
in the final signal region.  

The second function is designed to distinguish signal from 
CC events and is optimized, in particular, to reject 
$\nu_{\mu}$ CC interactions: 
$$
\likcc \define 
[[\; R_{Q_{T}}, R_{p_{T}}, \theta_{\nu l} \;], \evis, p_T^{m}, M_{T} \;]
$$
where the first part is similar to the function 
used for the muon tagging procedure (Section~\ref{sec:muveto}). 
The function \likcc uses variables referring to 
{\em the tagged muon $\lmucc$} and therefore,  
for events where the lepton is correctly tagged, 
it represents the actual CC kinematics (Figure~\ref{fig:mucckine}).  
The potentially different choice of leading particle(s) 
provides, globally, four additional 
degrees of freedom (including the $R_{Q_{T}}$ variable) 
as explained in Section~\ref{sec:sele}, and thus further 
justifies the use of two distinct likelihood functions. 
The correlations between some kinematic variables included 
in $\likcc$ are shown in Figure~\ref{fig:tau_vars} (right plots). 
For each event, a likelihood ratio $\ratcc$  
is computed as the ratio of the $\likcc$ functions constructed 
from the selected $\lmucc$ in signal events and from the 
true muon in MC $\nu_{\mu}$ CC interactions passing the FLV. 

Events are plotted in the plane $[\ratnc,\ratcc]$, 
as shown in Figure~\ref{fig:like2D} 
for simulated signal ($0\gamma$) and backgrounds. Two 
distinct populations, corresponding to the NC and CC backgrounds,  
are clearly visible, demonstrating  
that indeed the two background sources 
are independent and have opposite overall kinematics. 
The signal region (the blind box), further subdivided into 
different bins, lies at large values of both  
likelihood ratios. This corresponds to the selection of 
events where the \xvis candidate is isolated from the 
hadronic system and is not fully balanced in the transverse plane.  

For all topologies the analysis reaches its optimal sensitivity  
at low background levels (Table~\ref{tab:lowbkg}). This is 
an indication that the kinematic rejection of background is optimal. 
The sensitivity curves of Figure~\ref{fig:sensiall} 
clearly show a fast rise after the minimum   
at large values of the likelihood ratios,  
associated to regions of decreasing signal efficiency and 
almost background-free. The preliminary quality 
cuts on the statistical significance of the main  
variables (Section~\ref{sec:principles}) allows us  
to exploit such extreme values of the likelihood ratios.  
Due to the efficient kinematic suppression of the main 
background sources (NC and \numu CC), the very small 
residual background in these regions consists almost 
entirely of \nue(\anue) CC events.  

The choice of the binning for the signal region is 
performed {\em before} analyzing data events and is  
the one providing the best overall sensitivity to 
oscillations (Figure~\ref{fig:sensiall}).  
This criterion allows an objective definition of bins. 
First, the highest bin is chosen such that its lower edge 
is as close as possible to the minimum of the curves 
of Figure~\ref{fig:sensiall}, while consistent with a 
background of less than 0.5 events. 
Additional bins are then included in the signal region,  
up to the point where no further significant improvement 
of the combined sensitivity is obtained.  
The final binning follows the scheme of Figure~\ref{fig:like2D} 
for $0\gamma$ and $1\gamma$ events. In the $2\gamma$ 
topology, due to the limited statistics, 
only \ratnc is used to define the binning.  
For the $3h$ topology, the signal bins use also 
$\ratint$ as a third dimension, since the higher multiplicity 
enhances the effect of the internal \xvis 
structure on background rejection. The last bin 
along \ratnc is thus further divided into three bins  
along \ratint and \ratcc respectively. 

Table~\ref{tab:hadrons} summarizes the analysis flow on the 
various samples. The signal region is defined as  
$\ratcc > 3.5,0,0,0$ and $\ratnc > 5.5,6.0,5.5,4.0$ 
for the $0\gamma$, $1\gamma$, $2\gamma$ and $3h$ topologies 
respectively. The absence of a \xvis internal structure 
necessitates a tighter constraint on \ratcc for the $0\gamma$ 
topology. In the $3h$ decays, due to the large $a_{1}$ mass, 
a smaller phase space is available for the final state 
$\nu_{\tau}$, thus increasing the effectiveness of the \ratnc 
cut. The final signal efficiency is similar for all 
topologies (Table~\ref{tab:hadrons}).  
All signal bins are listed in Table~\ref{tab:allbins}.

\begin{table}[tb]
\caption{
Number of background and data events in the signal region. 
The corresponding $\ntaumu$ and $\ntaue$, as defined in
Sections~\ref{sec:systau} and~\ref{sec:limits}, are 
listed in the last two columns.  
The $1/2\gamma$ and $0/1$-$2\gamma$ topologies contain overlap events. 
The bins denoted by a star are considered as low background 
bins (Section~\ref{sec:allcha}).  
}
\label{tab:allbins}
\vspace*{0.2cm}
\begin{center}
\begin{tabular}{lcccccccc}
  \hline
\multicolumn{3}{c}{Analysis} & Bin \# & Tot bkgnd & Data & $\ntaumu$ & $\ntaue$  & \\ \hline 
$\nu_{\tau}h(n\pi^0)$ & DIS 
& $0\gamma$ & I & $4.49 \pm 1.50$ &  5 & 454 & 10.4  & \\
& & $0\gamma$ & II & $3.07 \pm 1.17$ &  5 & 345 & 8.2  & \\
& & $0\gamma$ & III & $0.05\,^{+\,0.60}_{-\,0.03}$ &  0 & 288 & 6.9 & * \\
& & $0\gamma$ & IV & $0.12\,^{+\,0.60}_{-\,0.05}$&  0 & 1345 & 31.1 & * \\ \cline{3-9} 
& & $1\gamma$ & I & $4.47 \pm 1.58$ &  5 & 283 & 6.8  & \\
& & $1\gamma$ & II & $1.54\pm0.89$&  0 & 244 & 5.7 & \\
& & $1\gamma$ & III & $0.07\,^{+\,0.70}_{-\,0.04}$ &  0 & 223 & 5.7  & * \\
& & $1\gamma$ & IV & $0.07\,^{+\,0.70}_{-\,0.04}$&  0 & 1113 & 26.6  & * \\ \cline{3-9} 
& & $2\gamma$ & I & $2.57 \pm 0.91$ &  3 & 318 & 7.4 & \\
& & $2\gamma$ & II & $0.66\pm0.44$ &  0 & 175 &  4.1 & \\
& & $2\gamma$ & III & $0.49 \pm 0.40$ &  0 & 82 & 1.9  & \\
& & $2\gamma$ & IV & $0.11\,^{+\,0.60}_{-\,0.06}$&  0 & 211 & 4.9  & * \\ \cline{3-9} 
& & $1/2\gamma$ & I & $1.40 \pm 0.77$ &  2 & 154 & 3.7  & \\
& & $1/2\gamma$ & II & $0.17\,^{+\,0.70}_{-\,0.08}$ &  0 & 124 & 2.9  & \\
& & $1/2\gamma$ & III & $0.20\,^{+\,0.70}_{-\,0.06}$ &  1 & 707 & 16.9  & * \\ 
& & $0/1$-$2\gamma$ & IV & $0.14\,^{+\,0.70}_{-\,0.06}$ &  0 & 1456 & 34.2  & * \\ \hline 
$\nu_{\tau} 3h(n\pi^0)$ & DIS 
& $3h$ & I & $2.61\pm0.99$&  2 & 170 & 4.0  & \\
& & $3h$ & II & $0.58 \pm 0.57$ &  0 & 139 & 3.4  & \\
& & $3h$ & III & $0.86\pm0.57$&  0 & 74 & 1.7  &  \\  
& & $3h$ & IV & $0.55 \pm 0.59$ &  1 & 309 & 7.6  & \\
& & $3h$ & V & $0.32\,^{+\,0.57}_{-\,0.32}$&  0 & 675 & 16.6  & * \\ \hline  
\end{tabular}
\end{center}
\end{table}

\section{Reliability of the background estimate} 
\label{sec:bkgrel} 

Since a $\tau$ signal would appear as a statistically significant 
excess of events inside the signal region, a crucial point for the 
analysis is the control of background predictions in this 
region. This requires two distinct steps. First, 
corrections to the Monte Carlo are extracted from the data themselves.  
Then, data control samples are used to check 
final predictions and to evaluate the corresponding 
systematic uncertainties.

\subsection{Data simulator corrections} 
\label{sec:corr}

The procedure to evaluate the backgrounds is based  
upon the data simulator method (Section~\ref{sec:ds}), which 
implies the application of the full selection scheme to appropriate 
data (DS) and simulated (MCS) samples. However, the 
overall background is actually composed of three different physical 
sources: NC, $\nu_{\mu}(\bar{\nu}_{\mu})$ CC and $\nu_{e}(\bar{\nu}_{e})$ 
CC interactions (Table~\ref{tab:hadrons}). 
In order to obtain a precise background evaluation, these  
three categories are considered separately: 
\begin{itemize} 
\item{ 
The NC correction is obtained from {\em identified} $\nu_{\mu}$ CC  
interactions, by the removal of the leading muon. This results 
in hadronic systems of different charge distribution than 
genuine NC events. This charge bias is corrected for by selecting 
\xvis of both positive and negative charge and by averaging the 
two results. 
}   

\item{  
The $\nu_{\mu}(\bar{\nu}_{\mu})$ CC correction takes into account two 
distinct effects: the muon identification efficiency (veto) and 
the actual kinematic selection. The first correction factor is measured 
from a large sample ($\sim 6\times10^{6}$) of muons originating from 
a nearby beam and crossing the NOMAD detector. The kinematic 
correction factor is evaluated from {\em identified} $\nu_{\mu}$ CC 
interactions in which the identification of the leading muon is 
ignored and, instead, the full $\tau$ selection is performed.  
The topological bias related to unidentified muons (muon chamber 
acceptance, low muon momenta) has little effect since this class 
of events is efficiently rejected by the kinematic veto 
(Section~\ref{sec:muveto}) and is checked through an appropriate 
CC control sample (Section~\ref{sec:control}).  
} 

\item{  
The $\nu_{e}(\bar{\nu}_{e})$ CC correction is obtained from 
events {\em identified} as $\nu_{e}$ CC in the 
$\tau \rightarrow e\bar{\nu}_{e}\nu_{\tau}$ DIS search~\cite{PLB1}. 
These events are then passed through the full event selection by 
ignoring the lepton identification. The electron identification 
(veto) used in the analysis, mainly based on TRD requirements,   
was independently checked using a sample of high-energy $\delta$ 
rays produced in the DC volume by muons originating from the 
nearby beam. As explained in Section~\ref{sec:elveto},  
unidentified $\nu_{e}(\bar{\nu}_{e})$ CC do not have specific topologies. 
Therefore, the described correction is adequate for 
all kinematic configurations.  
}  
\end{itemize}  
The total net correction factors, $\epsilon_{\rm DS}/\epsilon_{\rm MCS}$, 
to the number of background events  
computed from the Monte Carlo are 1.8 for $0\gamma$, 2.0 for $1\gamma$, 
1.3 for $2\gamma$ and 1.1 for $3h$ topologies respectively. 
These numbers are integrated over the whole signal region 
and are dominated by the data simulator corrections of the NC samples. 
In the most sensitive kinematic region, characterized by 
only a small residual CC background (Table~\ref{tab:lowbkg}), 
the corresponding correction factors are 
0.70 for $0\gamma$, 0.91 for $1\gamma$, 0.96 for $2\gamma$ 
and 0.99 for $3h$ topologies respectively.

The data simulator corrections computed for \nutau CC events 
are 0.86 for single prong and 0.99 for three prong events and 
are essentially independent of the values of the likelihood ratios.

\begin{sidewaysfigure}[p]
\bc
  \epsfig{file=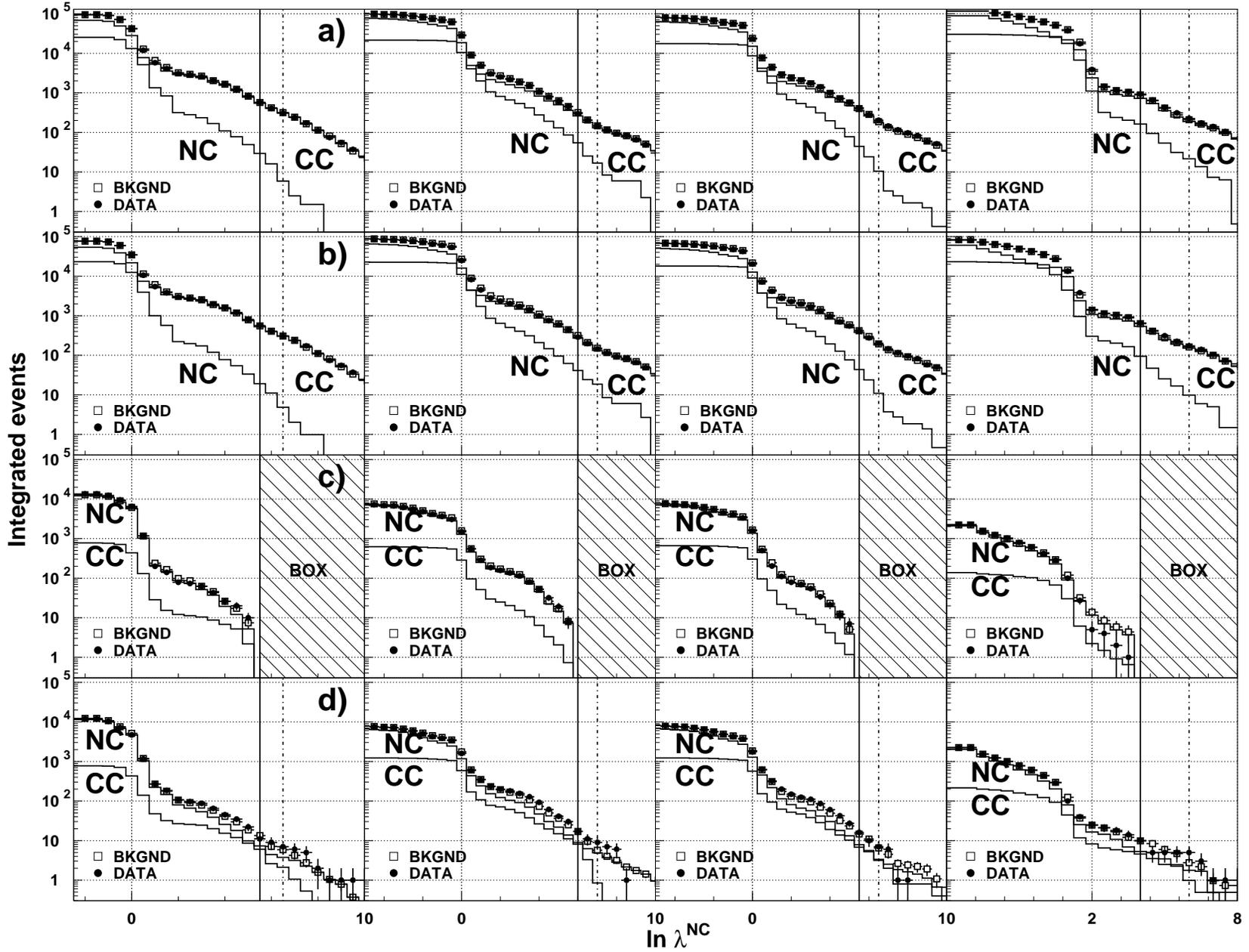,width=1.\textwidth}
\ec
\vspace*{-0.4cm}
\caption{
Cumulative $\ratnc$ distributions in control samples
for the $0\gamma$, $1\gamma$, $2\gamma$ and $3h$ topologies
(from left to right): a) after cuts on the structure 
of the hadronic system $H$ (Section~\ref{sec:jet});
b) CC control samples complementary to the signal region;
c) final distribution outside the signal region (difference
between the previous two); d) wrong sign analysis.
The histograms represent the individual NC and CC
contributions to the total background.
The vertical lines show the starting point of the signal region
(solid) and of the last signal bin (dashed).
The region to the right of the vertical solid lines in plots a)
and b) is used to estimate the systematic uncertainties on
the CC sample (region A of Figure~\ref{fig:like2D}).
}
\label{fig:allplots}
\end{sidewaysfigure}

\subsection{Control samples} 
\label{sec:control}

The definition of appropriate control samples, which are 
needed to validate the final background predictions,   
must take into account two problems.  
First, the possibility to independently check each of the 
individual background contributions is desirable and 
requires, in turn, the ability to discriminate among them.  
In addition, the statistical uncertainty associated with 
each control sample must be small in order to 
provide a significant test of systematic effects.   

The selection scheme based on two 
distinct likelihood functions provides by construction 
a separation between NC and CC back\-grounds.  
The\-re\-fo\-re, the required control samples can be defined 
in a natural way by selecting different regions 
in the plane of Figure~\ref{fig:like2D}. Since $\ratcc$ is built only 
from the $\nu_{\mu}$ CC sample, this latter background 
can be further partially decoupled from $\nu_{e}$ CC interactions  
in the plane of Figure~\ref{fig:like2D}. This is important since 
in the low-background region  
essentially only $\nu_{e}(\bar{\nu}_{e})$ CC events are present.  
However, for the sake of clarity, we will regroup backgrounds  
into NC and CC ($\nu_{\mu}$ and $\nu_{e}$) interactions in the 
following, unless otherwise specified.    

The CC control sample is defined by analysing the 
projection along $\ratnc$ of events which fail at least one of
the following selection criteria:
\begin{itemize}
\item{lepton veto;}
\item{$\ratint$ cut;}
\item{$\ratcc$ cut.}
\end{itemize}
At large values of $\ratnc$ this sample is completely dominated
by CC interactions, as can be seen from Figure~\ref{fig:allplots}a-b.

\begin{sidewaysfigure}[p]
\bc
  \epsfig{file=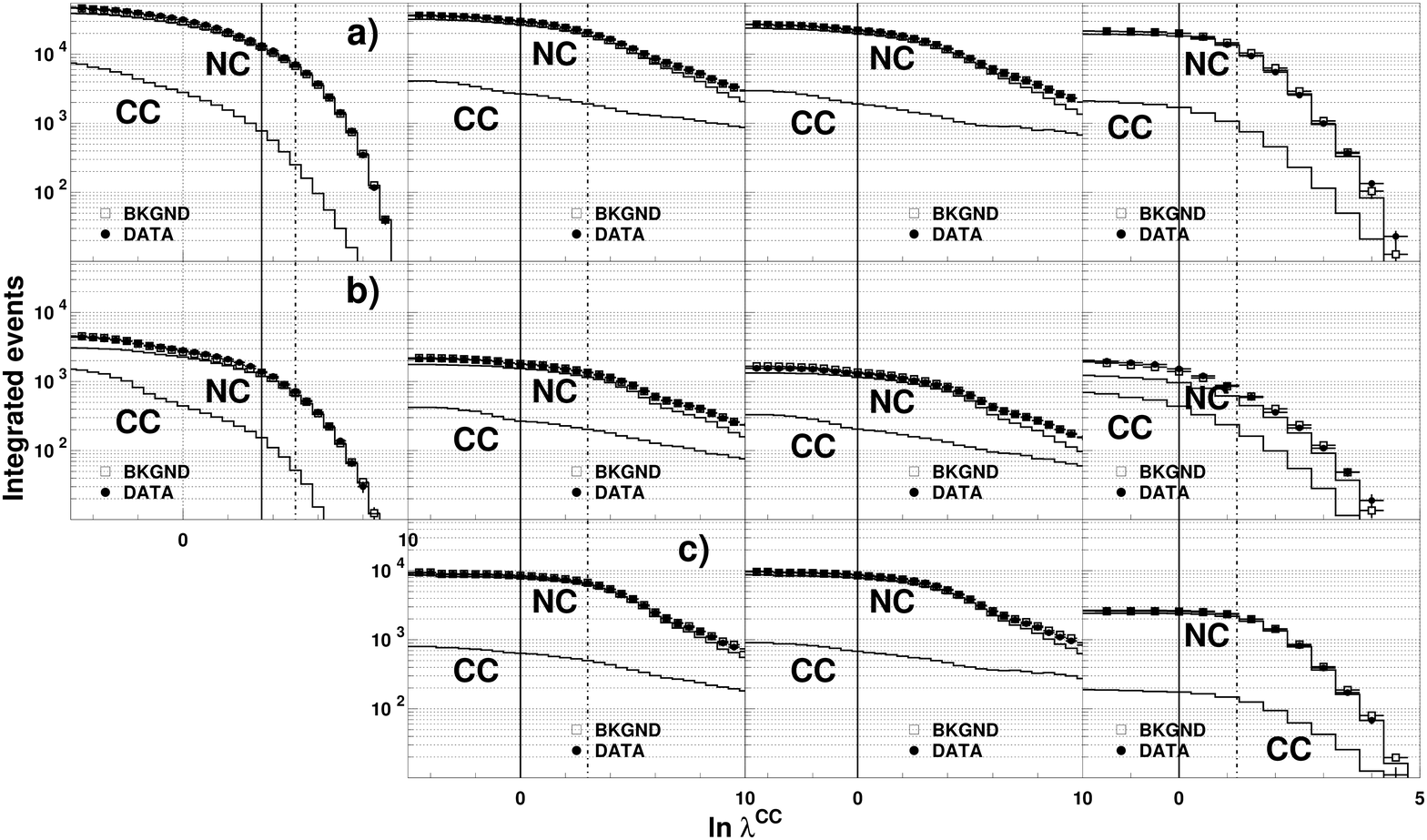,width=1.\textwidth}
\ec
\caption{
Cumulative $\ratcc$ distributions of data (full circles) and background
predictions (open squares) in NC control samples for the $0\gamma$, $1\gamma$,
$2\gamma$ and $3h$ (from left to right) topologies: a) without cuts on
\ratnc and \ratint; b) with a \ratnc cut reducing the overall
background by more than a factor of 10 but without \ratint cut;
c) with $\ratint>0.5,0.0,0.0$ (signal selection) but without
\ratnc cut. The histograms represent the
individual NC and CC contributions to the total background.
The vertical lines show the starting point of the signal region
(solid) and of the last signal bin (dashed).
The region to the right of the vertical solid lines is used
to estimate the systematic uncertainties on the NC sample
(region B of Figure~\ref{fig:like2D}).
}
\label{fig:ncsamp}
\end{sidewaysfigure}

Conversely, the NC control sample is defined by analysing the 
projection along $\ratcc$ of the events which fail the $\ratnc$ cut, 
after applying all the remaining cuts.  
Figure~\ref{fig:ncsamp} shows the statistical significance 
of the individual NC and CC contributions to this sample 
(see also Figure~\ref{fig:likint}).

In addition, the distributions of $\ratnc$ for the $\tau^{+}$ 
sample, where no signal is expected because of the small
$\anumu$ content of the beam (Section~\ref{sec:beam}),  
are compared with data for each of the individual steps 
of Table~\ref{tab:hadrons}. A similar check is performed in 
the $\tau^{-}$ search for the initial selection and for 
events outside the signal region (Figure~\ref{fig:allplots}).  
Due to the charge bias of the lepton tagging procedure 
(Section~\ref{sec:veto}) and, consequently, of kinematics 
(Section~\ref{sec:kine}), the $\tau^{+}$ selection is 
less effective than the $\tau^{-}$ search in rejecting backgrounds 
(Table~\ref{tab:hadrons}). This gives the possibility 
to check background predictions with larger statistics.   

Data events are in good agreement with background 
predictions for all control samples, as shown in 
Table~\ref{tab:hadrons}, Figure~\ref{fig:allplots}, 
Figure~\ref{fig:ncsamp} and Figure~\ref{fig:likint}. 
This gives confidence in the background estimation 
procedures and allows at the same time an evaluation 
of the systematic uncertainties.

\section{Systematic uncertainties} 
\label{sec:syserr}

The use of likelihood functions incorporating the full event 
topology provides a better estimate of systematic 
uncertainties with respect to a selection based on the application 
of a sequence of cuts. Moreover, the separation between CC and NC 
backgrounds gives additional high statistics control 
samples to constrain background predictions,  
as described in Section~\ref{sec:control}. 
This results, in turn, in a more precise background estimation.

\subsection{Background} 

The systematic uncertainties on the number of background 
events predicted inside the signal region can be divided 
into two contributions, related to the overall normalization and  
to the shape of the likelihood ratios.

\begin{figure}[tb]
\bc
  \hspace*{-0.2cm}\epsfig{file=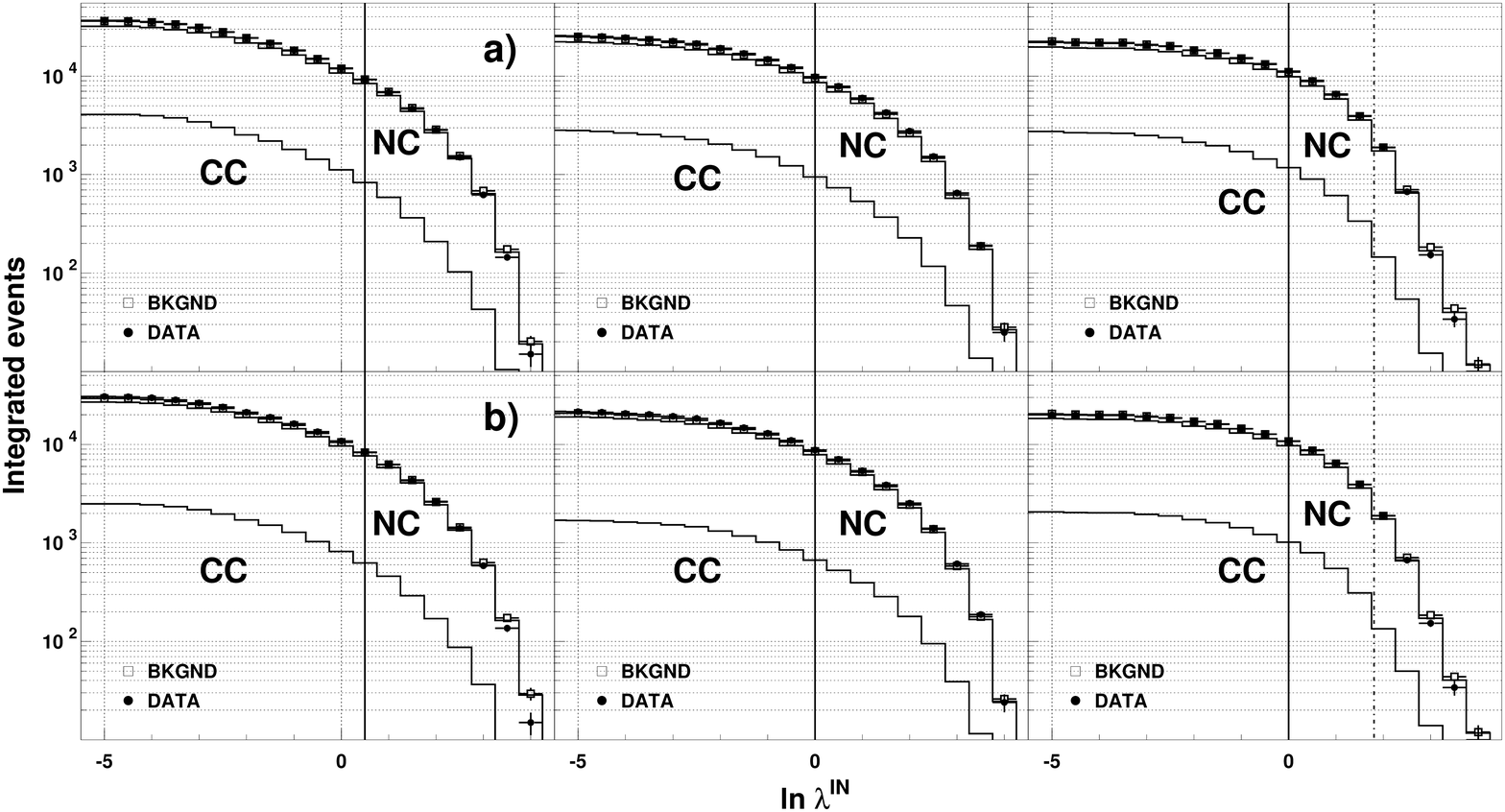,width=1.05\textwidth}
\ec
\caption{
Cumulative \ratint distributions of data (full circles) and background
predictions (open squares) for the $1\gamma$, $2\gamma$ and 
$3h$ (from left to right) topologies: a) without cut on  
\ratcc; b) with $\ratcc>0.0,0.0,0.0$ (signal selection).  
The histograms represent the individual NC and CC contributions 
to the total background. The vertical lines show the starting 
point of the signal region (solid) and of the last signal 
bin for $3h$ events (dashed).
The region to the right of the vertical solid lines is used
to estimate the systematic uncertainties. 
}
\label{fig:likint}
\end{figure}

The first term is estimated from a comparison of the total 
number of data and background events in both $\tau^{+}$ and 
$\tau^{-}$ searches as a function of the selection criteria. 
This results in a r.m.s. of 1.9\% from the differences in the 
last columns of Table~\ref{tab:hadrons}.  
This value includes uncertainties related to fluxes, cross-sections 
and to the effect of the selection cuts for \numu CC interactions,  
which define the normalization, and NC events, which are the  
dominant background sample. For the remaining components   
(different from \numu) there is an additional systematic uncertainty 
coming from the integral flux predictions~\cite{nutaubeam}. 
An upper limit on this contributions is obtained from the 
average systematic uncertainties on the corresponding energy 
spectra, which amount to 5.1\%, 8.5\% and 12.0\% for \nue, \anumu 
and \anue CC interactions respectively.  

The second term is evaluated from the two control regions  
shown in Figure~\ref{fig:like2D} for CC (region A) and  
NC (region B) backgrounds respectively. 
This is obtained by analyzing the corresponding control 
samples, defined in Section~\ref{sec:control} as the 
event projections along each of the two likelihood 
ratio axes (Figures~\ref{fig:ncsamp} and~\ref{fig:allplots}a-b).  
The systematic uncertainty is then estimated as the 
r.m.s. of the distribution of differences between data and 
predictions {\em inside the signal region} of each control sample.  
This procedure is based on the assumption that 
in the region A (region B) a \ratcc (\ratnc) cut does not 
introduce discrepancies between data and predictions 
which are dependent on the \ratnc (\ratcc) distribution. 
Due to the limited rejection factor of the \ratcc cut 
(Table~\ref{tab:hadrons}), this condition is fulfilled 
by the CC control sample and the resulting number (5.5\%) 
thus provides the final uncertainty on the CC background shape. 
 
The tighter constraint imposed on $\ratnc$ requires 
additional checks of the effect of this variable 
on the $\ratcc$ distributions in region B.  
This is achieved by comparing the level of the observed agreement 
with different cuts on $\ratnc$, chosen in such a way as to 
reduce the overall background by more than a factor of 10 with respect 
to the initial value (Figure~\ref{fig:ncsamp}a-b). 
Since the \xvis structure embedded in \ratint is 
almost independent of the global kinematics (apart 
from an overall boost factor in the laboratory frame), 
the cut on \ratint is not applied, in order to 
increase the available statistics and to 
have a consistent comparison of all topologies.  
The effect of the \ratint constraint is then checked 
separately, with no cut on \ratnc (Figure~\ref{fig:ncsamp}c). 
In addition, all the \ratint distributions  
for backgrounds are compared with data with and without 
the \ratcc cut (Figure~\ref{fig:likint}). 
The final systematic uncertainty (r.m.s.) 
inside the signal region is estimated to~be~4.1\% 
from all these control samples.  

Table~\ref{tab:systerr} summarizes the individual contributions 
to systematic uncertainties, which, added in quadrature, result 
in a total systematic uncertainty of 5.8\% for \numu CC, 
10.3\% for \anumu CC, 7.6\% for \nue CC, 13.3\% for \anue CC,  
and 4.5\% for NC respectively. However, 
the effect of this latter contribution is negligible since 
only the CC background is present in the most sensitive 
region of the analysis (Table~\ref{tab:lowbkg}). 
The overall net systematic 
uncertainty on the final background predictions  
for $0\gamma$, $1\gamma$, $2\gamma$ and $3h$ topologies  
is then 5.0\%, 5.1\%, 5.4\% and 5.5\% at the starting point   
of the signal region and 10.0\%, 9.2\%, 9.2\% and 5.8\% 
in the low background region.  
Given the large rejection factors of  
$\mycal{O}(10^{5})$ such a control of systematic 
uncertainties is noteworthy.

\begin{table}[tb]
\caption{
Contributions to systematic uncertainties for background and signal events. 
}
\label{tab:systerr}
\vspace*{0.2cm}
\begin{center}
\begin{tabular}{lccr}  \hline 
\multicolumn{4}{c}{Background} \\  \hline 
Normalization  &&&  1.9\% \\ 
Integral \nue/\numu &&& 5.1\% \\ 
Integral \anumu/\numu &&& 8.5\% \\ 
Integral \anue/\numu &&& 12.0\% \\ 
CC $\ln \lambda$ shape in signal region   &&& 5.5\% \\ 
NC $\ln \lambda$ shape in signal region   &&& 4.1\% \\ \hline  
\multicolumn{4}{c}{Signal}  \\ \hline 
Normalization  &&&  1.9\% \\ 
$\sigma_{\tau}/\sigma_{\mu}$   &&&  3.0\% \\ 
$\ln \lambda$ shape in signal region  &&&   5.0\% \\ \hline 
\end{tabular} 
\ec 
\end{table}

\subsection{Signal} 
\label{sec:systau}

The probability of oscillation, $P_{\rm osc}$, is estimated 
as the ratio between the number of observed $\tau$ events and the 
maximal number of signal events expected if all incident 
\numu had converted into \nutau (Section~\ref{sec:limits}). 
For the \numunutau oscillation, neglecting a term 
$\mathcal{O}(P_{\rm osc}^{2})$, this last quantity is defined by: 
\begin{equation}
\label{eqn:ntaumax}
\ntaumu = \Nmuobs \times (\epsilon_\tau/\epsilon_\mu)
          \times(\sigma_\tau /\sigma_\mu) \times Br
\end{equation}
where:
\begin{itemize}
\item $\Nmuobs$ is the observed number of $\numu$ CC interactions
(Section~\ref{sec:intro}). The number of $\numu$ CC interactions 
corresponding to the LM topologies are evaluated  
to be 11\% of the total~\cite{PLB1}.

\item $\epstau$ and $\epsmu$ are the detection efficiencies 
for $\tau$ signal events and $\numu$ CC events respectively, 
integrated over the incident \numu spectrum.  
The cuts used to select $\Nmuobs$ and $\epsmu$ 
vary from channel to channel in order
to reduce systematic uncertainties in the ratio
$\epstau/\epsmu$ for that channel.  

\item $\sigma_\tau /\sigma_\mu$ is the suppression factor of the
$\nutau$ cross section due to the difference between the $\tau$
and $\mu$ masses, averaged over the incident \numu spectrum. 
For the $\nu_{\mu}$ spectrum used in this 
experiment~\cite{nutaubeam} and for an energy-independent oscillation 
probability (corresponding to the large $\Delta m^{2}$ hypothesis),
it is evaluated to be 0.48, 0.60 and 0.82
for the deep inelastic, resonance and quasi-elastic processes.
The resulting average values for the DIS and LM analyses are
0.48 and 0.57 respectively.
The event sample selected by the LM analyses contains significant 
fractions of DIS events satisfying the LM selection~\cite{PLB1}. 

\item $Br$ is the branching ratio for the $\tau$ 
decay channel under consideration.
\end{itemize}

The systematic uncertainty on the overall normalization for 
the signal sample is common to the \numu CC background (1.9\%) and 
includes the effect of the individual selection criteria listed in 
Table~\ref{tab:hadrons}. The maximum Data Simulator correction 
applied to the signal likelihood shapes, decoupled from the 
overall normalization, is 5.0\%. We use this maximum 
correction as the systematic uncertainty on the signal shape.  

The uncertainty on the suppression factor $\sigma_{\tau}/\sigma_{\mu}$, 
which will also contribute to the overall systematic 
uncertainty, is due to the uncertainties on the $\nu_{\mu}$ 
energy spectrum and on the structure functions used in the 
computation. It is estimated to be~3.0\%. The uncertainty 
on the $\tau$ branching ratios is negligible~\cite{PDG}. 

The final systematic uncertainty on the signal sample 
is~6.1\% from the sum in quadrature of all the individual 
contributions summarized in Table~\ref{tab:systerr}.

\section{Results} 
\label{sec:results}

\subsection{Analysis of the signal region} 
\label{sec:boxopen} 

\begin{sidewaysfigure}[p]
\bc
  \rotatebox{0}{\begin{tabular}{cc} 
  \epsfig{width=8.5cm,file=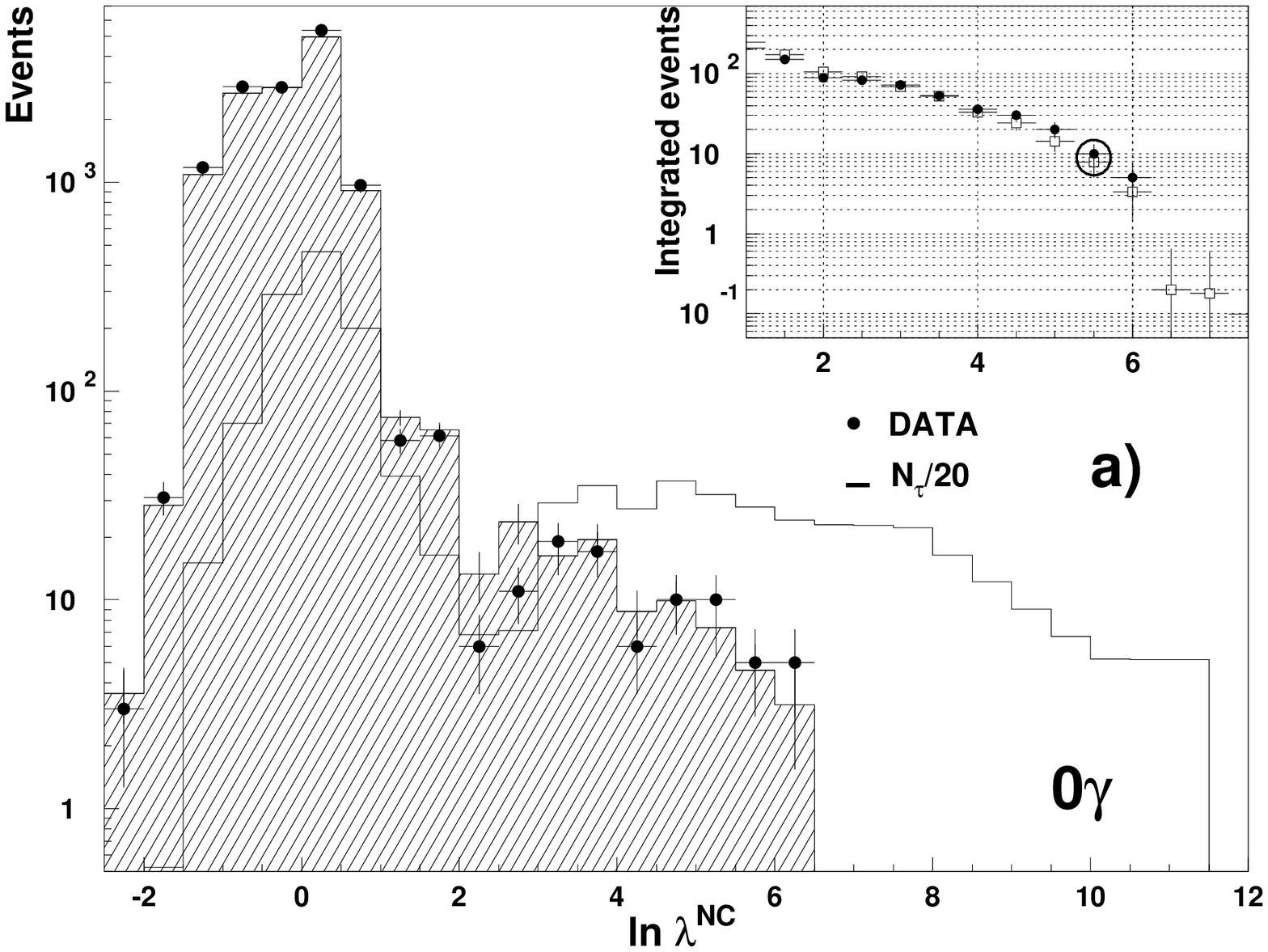}  & 
  \epsfig{width=8.5cm,file=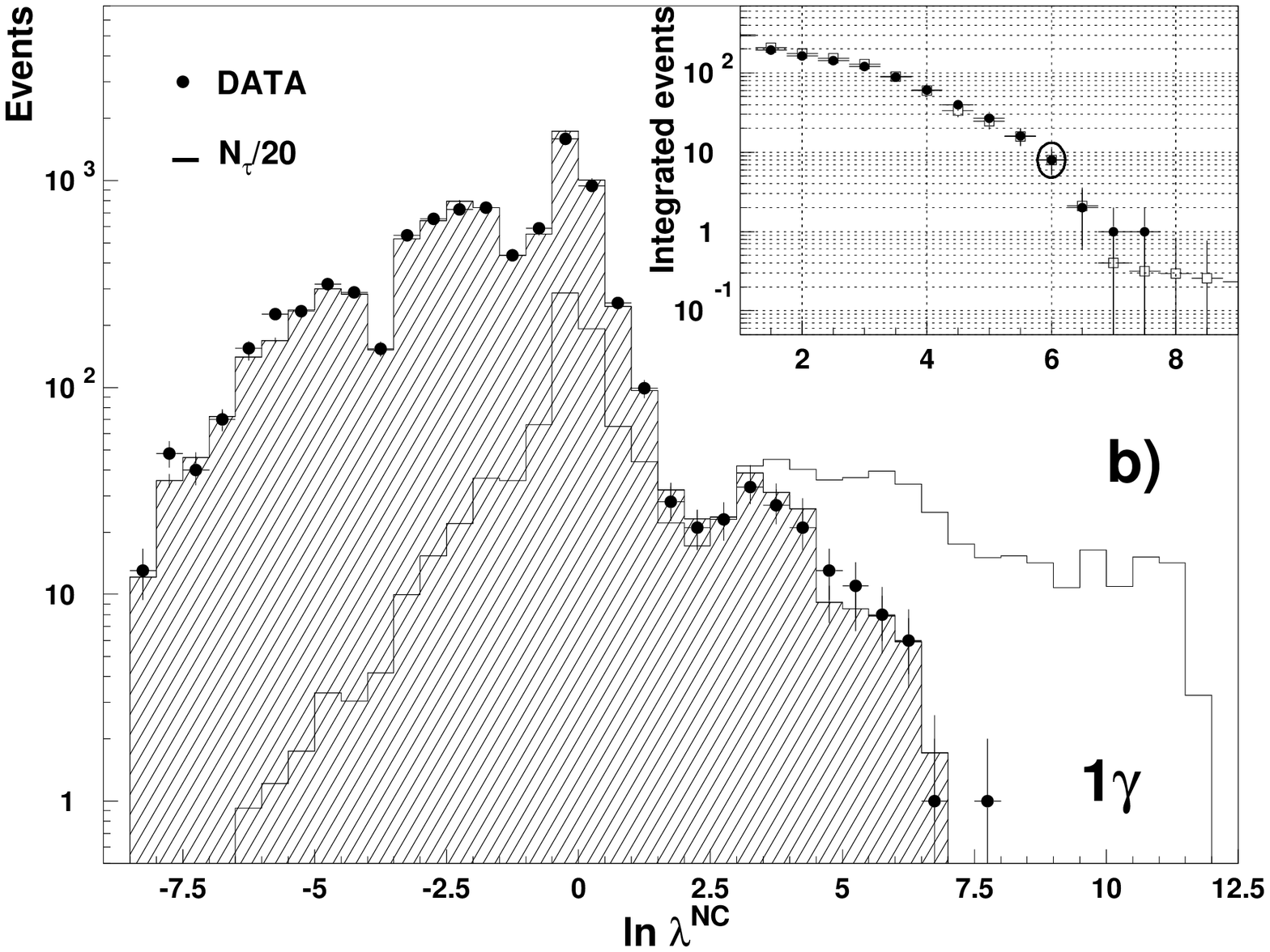} \\ 
  \epsfig{width=8.5cm,file=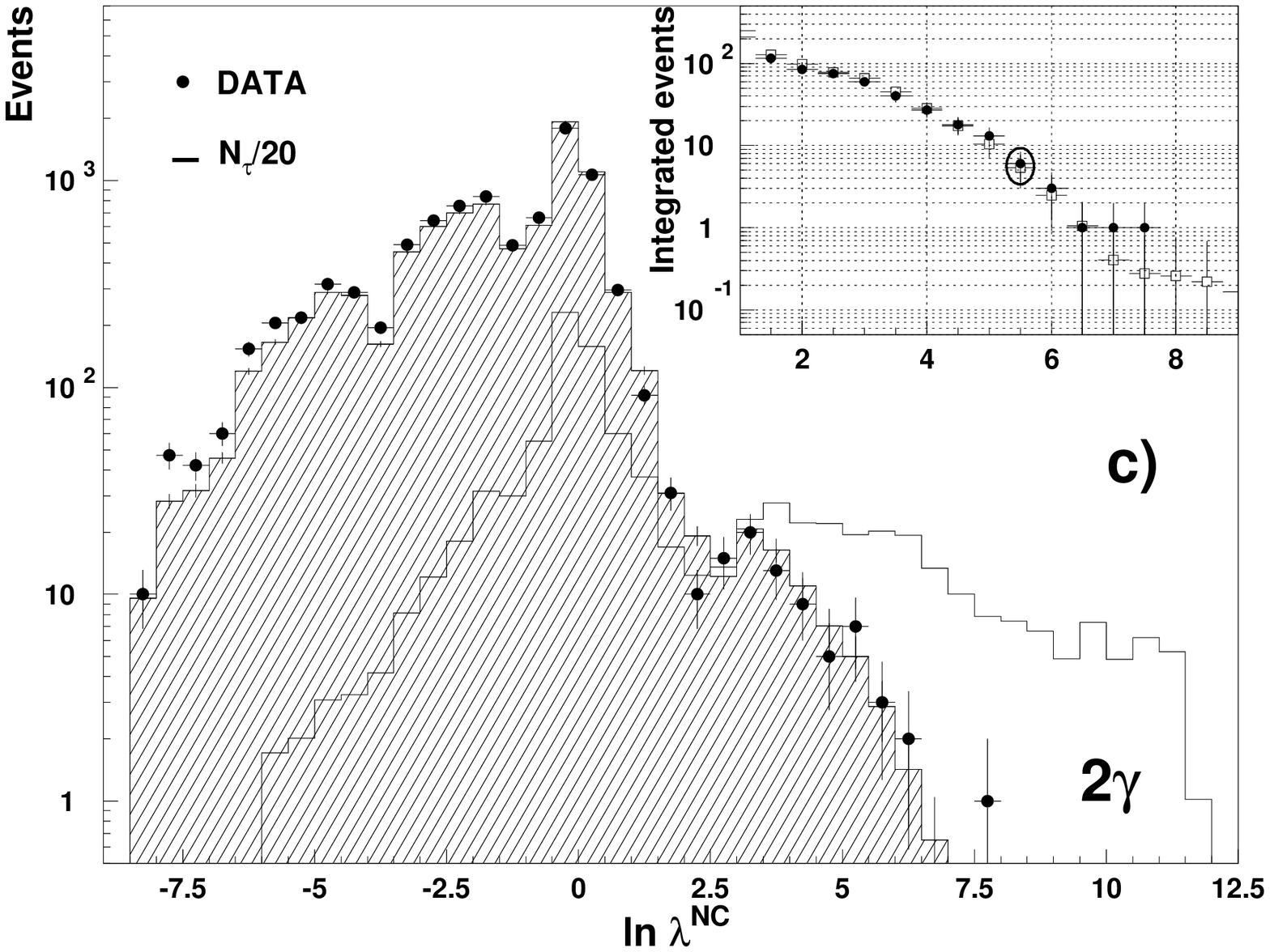} &  
  \epsfig{width=8.5cm,file=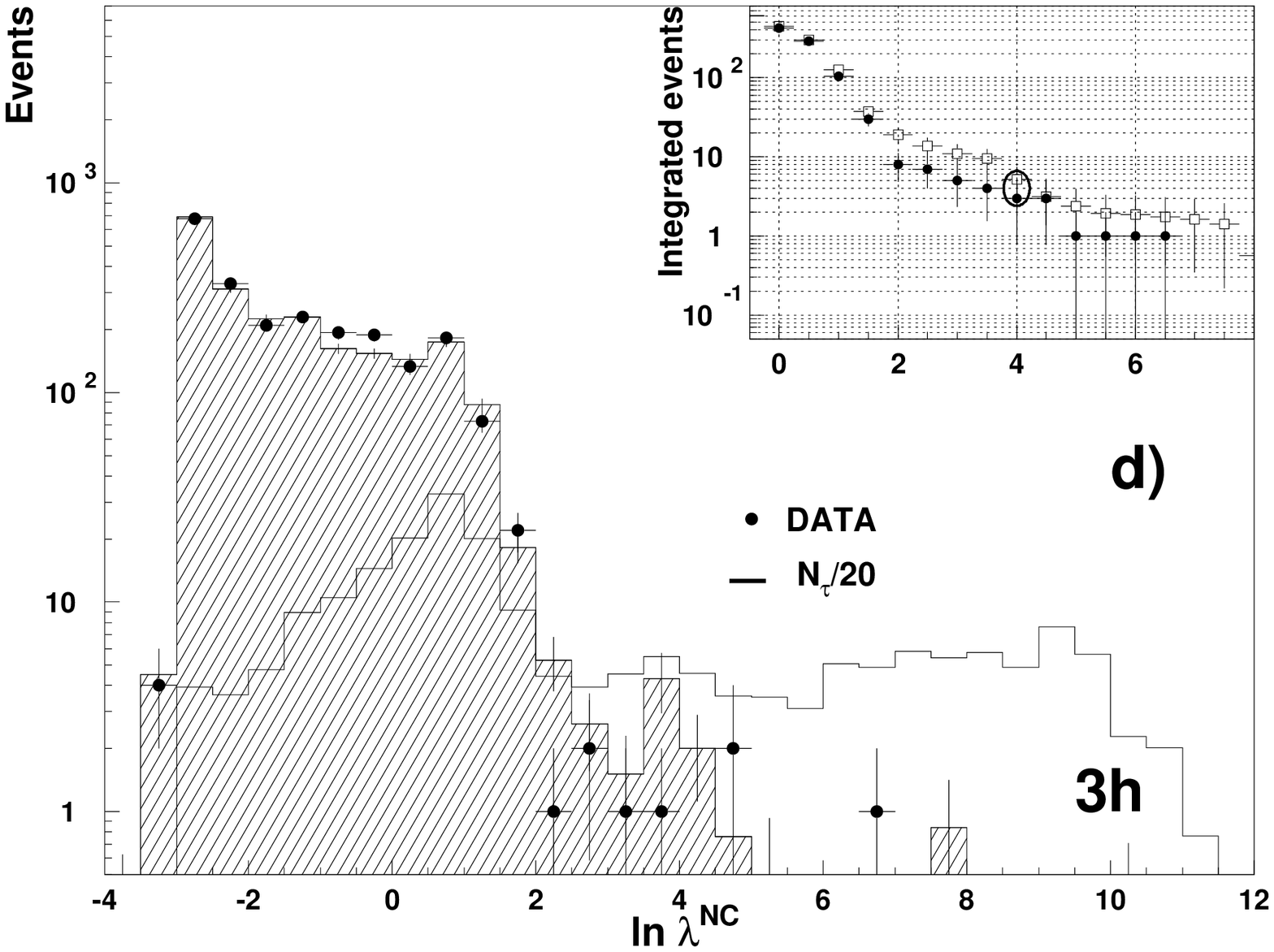} \\ 
  \end{tabular}} 
\ec
\caption{ 
Histograms of $\ratnc$ for events passing all cuts except the 
cut on $\ratnc$ in a) $0\gamma$, b) $1\gamma$, c) $2\gamma$ 
and d) $3h$, for data (points with statistical error bars), 
total backgrounds (shaded) and $\ntaumu$ (Section~\ref{sec:systau}),  
scaled down by a factor 20 (unshaded).
The insets give, for each value of $\ratnc$, the total number of
events beyond that value, for data (dots) and expected background (squares);
the encircled points are at the boundary of the signal regions.
}
\label{fig:finalbkg}
\end{sidewaysfigure}

After the choice of the selection criteria and the cross-check 
of the background predictions, we analyze data events falling in the 
signal region of the hadronic DIS channels (the blind box). 
Data events populate the various bins in a manner consistent 
with backgrounds, as summarized in Table~\ref{tab:allbins}. 
The overall integrals and shapes of the final $\ln \lambda$ 
distributions are in good agreement with 
background predictions (Figure~\ref{fig:finalbkg}).  
Therefore, no signal from oscillations is observed.

\begin{table}[p]
\vspace*{-0.60cm}
\caption{
Summary of backgrounds and efficiencies for all 
the individual $\tau$ searches.   
The columns labelled $\tau^-$ summarize the observed number of $\tau^-$
candidate events (Obs.) and the corresponding predicted background 
(Tot Bkgnd) for the sum of all bins in the signal region. 
The columns labelled $\tau^+$ contain the equivalent numbers for the 
positive control sample. The corresponding $\tau^-$ selection 
efficiencies ($\epstau$), not including branching ratios, and 
the $\ntaumu$ and $\ntaue$ (Sections~\ref{sec:systau} 
and~\ref{sec:limits}) are also listed.
For the LM topologies the quoted $\epstau$ is the average efficiency 
for quasi-elastic and resonance events.
The last column shows the overall sensitivity of each 
individual $\tau$ search, based on the combination of all 
signal bins.   
}
\label{tab:CHANNELS}
\vspace*{0.2cm}
\begin{center}
\hspace*{-1.0cm}
\begin{tabular}{lccccccccccc}
\hline 
& & \multicolumn{2}{c}{\bf $\tau^{-}$} & & \multicolumn{2}{c}{\bf $\tau^{+}$}
 & & & & $S_{\mu\tau}$ \\ \cline{3-4} \cline{6-7}
 \multicolumn{2}{c}{Analysis}   & Obs. & Tot Bkgnd    &  & Obs. & Tot Bkgnd  
& $\epstau(\%)$ & $\ntaumu$ & $\ntaue$ & $(\times10^{-4})$ \\ \hline 
 $\nu_{\tau}\bar{\nu}_{e}e$ & DIS   & 5    & $5.3\,^{+\,0.7}_{-\,0.5}$ & & 9    & $8.0 \pm 2.4$ & 3.6  & 4318 & 88.0 & $8.0$  \\
$\nu_{\tau} h(n\pi^0)$  & DIS & 21  & $19.5 \pm 3.5$ & & 44  & $ 44.9 \pm 4.6$ & 2.2 & 7522 & 177.4 & $4.0$ \\ 
 $\nu_{\tau} 3h(n\pi^0)$ & DIS & 3  & $4.9 \pm 1.5$ & & 10 & $9.9 \pm 1.6$ &
 1.3 & 1367  & 33.3 & $22.2$ \\ 
 $\nu_{\tau}\bar{\nu}_{e}e$ & LM & 6 & $5.4 \pm 0.9$ & & 3 & $2.2 \pm 0.5$ & 6.3 & 864 & 8.8 & $55.2$  \\ 
$\nu_{\tau} h(n\pi^0)$ & LM & 12 & $11.9 \pm 2.9$ & & 40 & $44.1 \pm 9.2$ & 1.9 & 857 & 16.7 & $88.9$ \\ 
 $\nu_{\tau} 3h(n\pi^0)$ & LM & 5 & $3.5\pm1.2$ & & 1 & $2.2 \pm 1.1$ & 2.0  & 298 & 5.2 & $161.0$ \\
\hline 
\end{tabular}
\end{center}
\end{table}

\begin{table}[p] 
\caption{
Summary of background and data events in the low 
background bins. The corresponding $\ntaumu$ and $\ntaue$, as defined in
Sections~\ref{sec:systau} and~\ref{sec:limits}, are listed 
in the last two columns. The quoted background consists mainly 
of $\nu_{e}$ CC events for the $\nu_{\tau}e\bar{\nu}_{e}$ channel; 
of a mixture of $\nu_{e}$ and $\bar{\nu}_{e}$ CC events for the  
$\nu_{\tau}h(n\pi^0)$ channel and of $\nu_{\mu}$ CC events for the 
$\nu_{\tau} 3h(n\pi^0)$ channel. The errors include  
contributions from the remaining sources. 
}
\label{tab:lowbkg}
\vspace*{0.2cm}
\begin{center}
\begin{tabular}{lccccccc}
  \hline
\multicolumn{3}{c}{Analysis} & Bin \# & Tot bkgnd & Data & $\ntaumu$ & $\ntaue$  \\ \hline 
$\nu_{\tau}e\bar{\nu}_{e}$ & DIS & & III & $0.18\,^{+\,0.18}_{-\,0.08}$ &  0 & 680 & 15.0 \\
 & & & VI & $0.16\pm0.08$ &  0 & 1481 & 32.7 \\
\multicolumn{3}{c}{($\evis<12$ GeV)} & II+III+VI & $0.27\pm0.13$ &  0 & 665 & 8.7 \\ \hline 
$\nu_{\tau}h(n\pi^0)$ & DIS & $0\gamma$ & III & $0.05\,^{+\,0.60}_{-\,0.03}$ &  0 & 288 & 6.9 \\
& & $0\gamma$ & IV & $0.12\,^{+\,0.60}_{-\,0.05}$&  0 & 1345 & 31.1 \\  
& & $1\gamma$ & III & $0.07\,^{+\,0.70}_{-\,0.04}$ &  0 & 223 & 5.7 \\
& & $1\gamma$ & IV & $0.07\,^{+\,0.70}_{-\,0.04}$&  0 & 1113 & 26.6  \\  
& & $2\gamma$ & IV & $0.11\,^{+\,0.60}_{-\,0.06}$&  0 & 211 & 4.9  \\  
& & $1/2\gamma$ & III & $0.20\,^{+\,0.70}_{-\,0.06}$ &  1 & 707 & 16.9  \\  
& & $0/1$-$2\gamma$ & IV & $0.14\,^{+\,0.70}_{-\,0.06}$ &  0 & 1456 & 34.2  \\ \hline 
$\nu_{\tau} 3h(n\pi^0)$ & DIS & $3h$ & V & $0.32\,^{+\,0.57}_{-\,0.32}$&  0 & 675 & 16.6  \\ \hline\hline 
Total & & & & $1.69\,^{+\,1.85}_{-\,0.39}$&  1 & 8844 & 199.3  \\ \hline 
\end{tabular}
\vspace*{-0.60cm}
\end{center}
\end{table}

\begin{figure}[tb]
  \mbox{\epsfig{file=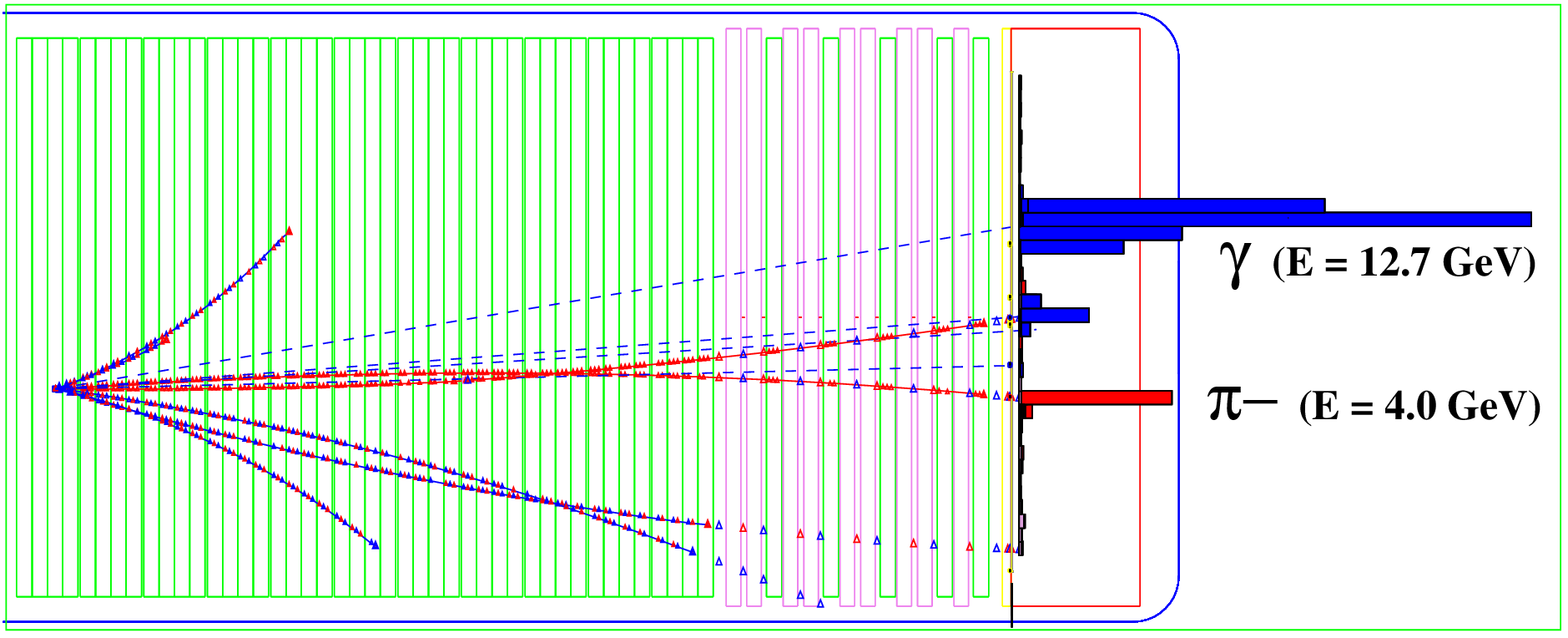,angle=-90,width=1.00\textwidth}} 
\vspace*{0.2cm} 
\caption{
Data event (run 14162, event 25050) falling in the low background 
signal region of Table~\ref{tab:lowbkg}. The event is classified 
as $\tau \rightarrow \rho \nu_{\tau}$ (bin $1/2 \gamma$ III) 
decay candidate: $\ratint=2.1$, $\ratcc=9.4$ and $\ratnc=7.5$. 
Solid lines represent the reconstructed charged tracks 
(open triangles show their extrapolation) whereas 
dashed lines represent the ECAL neutral clusters. The energy 
deposition of individual ECAL cells is also shown by 
the shaded bins on the right. 
}
\label{fig:taucand}
\end{figure}

\subsection{Combined results} 
\label{sec:allcha}

The overall NOMAD results are obtained from the combination of 
the new analysis of the hadronic $\tau$ decays with the 
$\nu_{\tau}e\bar{\nu}_{e}$ DIS and the LM analyses described 
in Ref.~\cite{PLB1} (Appendix~\ref{sec:others}). 
Table~\ref{tab:CHANNELS} lists all the  
individual contributions and the corresponding sensitivities. 
The unified analysis of $\nu_{\tau}h(n\pi^0)$ topologies described 
in the present paper is the most sensitive \nutau 
appearance search in NOMAD.  

As discussed in Section~\ref{sec:kine}, the strength of this   
search lies in the possibility to define, through kinematic 
constraints, large regions characterized by the expectation of 
low background. A precise control of background predictions 
using the data themselves is also crucial. The major low background 
($< 0.5$ events) bins are summarized in Table~\ref{tab:lowbkg}.  
Overall, these bins contribute about 75\% of the 
total sensitivity of the experiment. A single event 
is observed within this kinematic region.  
This event, shown in Figure~\ref{fig:taucand}, is classified 
as a $\tau \rightarrow \nu_{\tau}\rho$ candidate and has   
large values of all likelihood ratios (Figure~\ref{fig:finalbkg}b-c). 

The overall systematic uncertainties for the hadronic DIS 
channels are given in Section~\ref{sec:syserr}. For the remaining 
modes the estimated systematic uncertainties are 20\% and 10\% on 
backgrounds and \ntaumu respectively~\cite{PLB1}.

\subsection{Evaluation of confidence regions} 
\label{sec:limits}

The final result of the measurement is expressed as a frequentist
confidence interval~\cite{STAT} which accounts for the fact that each
$\tau$ decay mode and signal bin (Tables~\ref{tab:allbins} 
and~\ref{tab:others}) may have a different signal to
background ratio.\ The acceptance region of Ref.~\cite{STAT} therefore 
becomes multi-dimensional to contain each of the separate measurements.\  
The procedure follows the prescription of Ref.~\cite{PDG}. 
This computation~\cite{PLB2} takes into account the number of 
observed signal events, the expected background and its uncertainty, 
and the maximal number of the expected signal events.

\begin{sidewaysfigure}[p]
\hspace*{-0.8cm}
\bc 
\epsfig{file=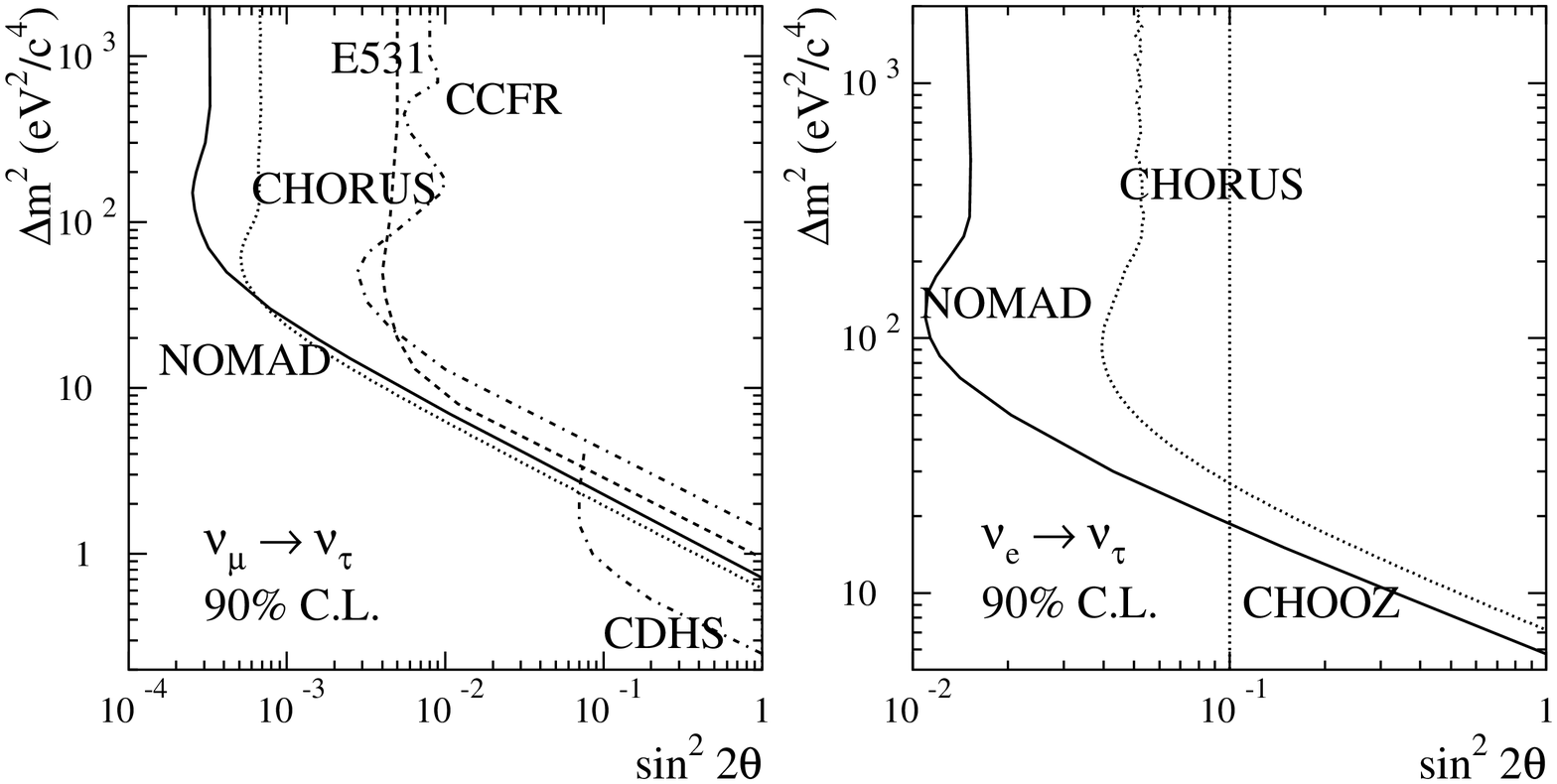,width=0.90\textwidth}
\ec 
\caption{
Contours outlining a 90 \% CL region in the $\Delta m^2 - \sin^22\theta$ 
plane for the two-family oscillation scenario. The NOMAD $\numunutau$ 
(left) and $\nuenutau$ (right) curves are shown 
as solid lines, together with the limits 
published by other experiments \protect\cite{E531,CDHS,CCFR,CHORUS,CHOOZ} 
} 
\label{fig:limit}
\end{sidewaysfigure}

\begin{figure}[tb]
\hspace*{0.4cm}
\bc 
\epsfig{file=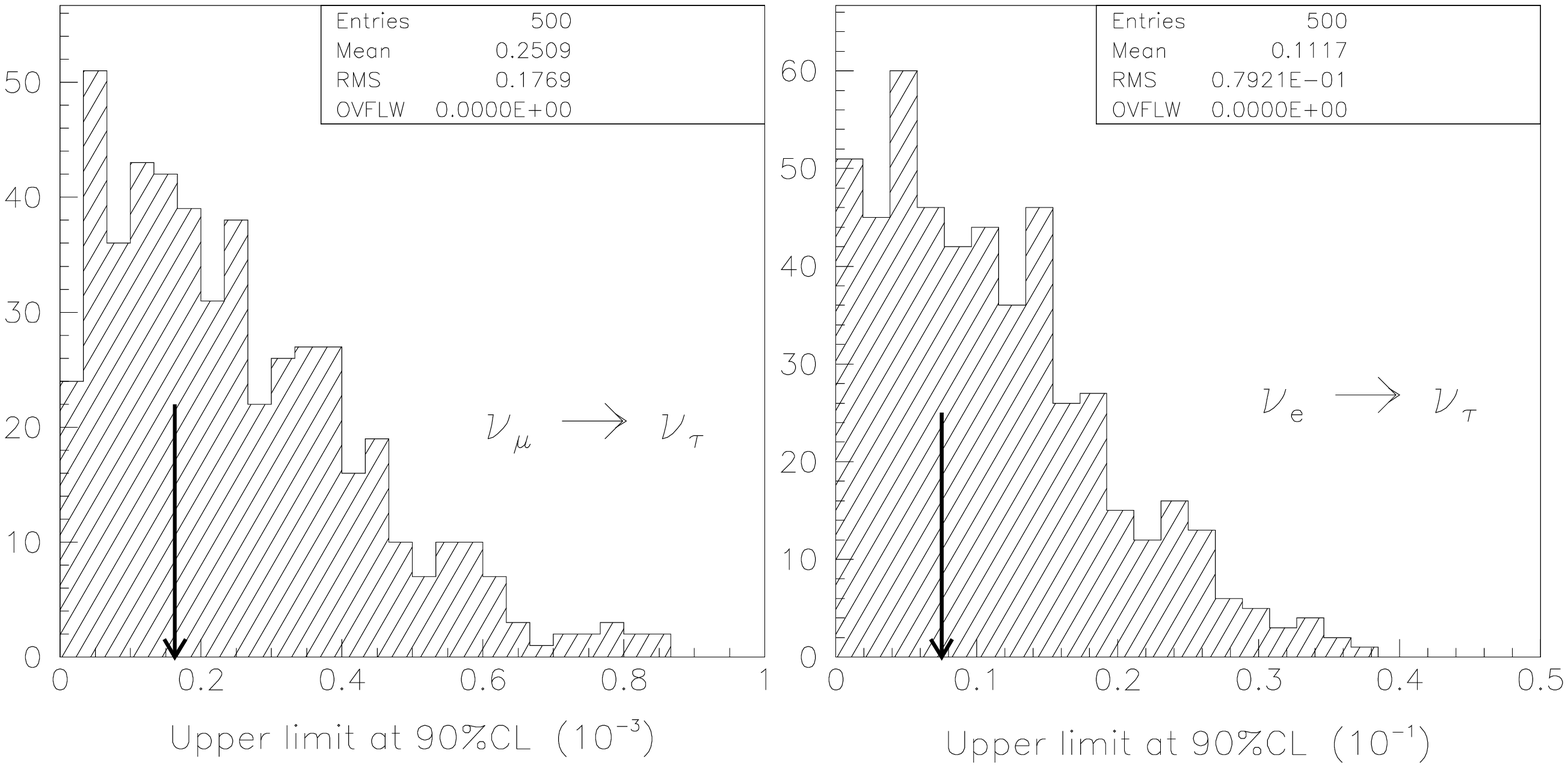,width=1.00\textwidth}
\ec 
\caption{Histograms of the upper limits obtained, in the absence of signal 
events, for 500 simulated experiments with the same NOMAD 
expected background \cite{STAT}. 
The averages correspond to the quoted sensitivities while the arrows 
are the actual upper limits obtained from the data in the $\numunutau$ 
(left) and $\nuenutau$ (right) searches. 
}
\label{fig:sens}
\end{figure}

The resulting 90\% C.L. upper limit on the $\numunutau$ 
oscillation probability is: 
\begin{equation}
P_{\rm osc} (\numunutau) <  1.63 \times 10^{-4}
\end{equation}
Under a two-neutrino family formalism this corresponds 
to $\sin^2 2\theta_{\mu \tau} < 3.3 \times 10^{-4}$ 
for large $\Delta m^2$ and to the exclusion region in the $\Delta
m^2 - \sin^2 2\theta$ plane shown in Figure~\ref{fig:limit}.
The result is significantly more stringent than the previously published  
limits \cite{PLB1,E531,CDHS,CCFR,CHORUS}. 
The sensitivity \cite{STAT} of the 
experiment is $P_{\rm osc} = 2.5 \times 10^{-4}$; this is higher than 
the quoted confidence limit, since the number of observed events is 
smaller than the estimated background. In the absence of signal events,
the probability to obtain an upper limit of $1.63 \times 10^{-4}$
or lower is 37\% (Figure \ref{fig:sens}). This result matches the 
design sensitivity of the experiment ($P_{\rm osc}=1.9\times 10^{-4}$).

In the context of a two-flavour approximation, we can reinterpret
the result in terms of $\nuenutau$ oscillations,
by assuming that any observed $\nu_{\tau}$ signal should 
be due to oscillations from the small $\nu_{e}$ 
component of the beam~\cite{PLB3}.
The corresponding maximal number of signal events,
$\ntaue$, is then obtained from $\ntaumu$ by reweighting the  
signal events using the $\nu_e$ to $\nu_{\mu}$ flux ratio.
This procedure introduces a further systematic uncertainty 
of 4.7\% on the values of \ntaue, related to flux 
predictions~\cite{nutaubeam}. The resulting 90\% C.L. upper 
limit on the $\nuenutau$ oscillation probability is then: 
\begin{equation}
\label{eqn:limit}
P_{\rm osc} (\nuenutau ) \; < \; 0.74 \times 10^{-2}
\end{equation}
corresponding to $\sin ^2 2 \theta_{e\tau} < 1.5 \times 10^{-2}$
for large $\Delta m^{2}$. The exclusion region in the
$\Delta m^{2} - \sin ^2 2 \theta$ plane is also shown in 
Figure \ref{fig:limit}. The $\nuenutau$ sensitivity is 
$P_{\rm osc} =  1.1 \times 10^{-2}$ and the probability to
obtain an upper limit of $0.74 \times 10^{-2}$ or lower is 39\% 
(Figure \ref{fig:sens}). Both the sensitivity and the 
probability to obtain the result (goodness-of-fit) are 
an essential part of the limits themselves, as pointed 
out in Ref.~\cite{PDG}\cite{STAT}. 

The results from the $\tau$ appearance search also exclude 
effective couplings of \numu or \nue with the $\tau$ 
lepton, which are equivalent to the oscillation 
probabilities at large $\Delta m^2$.  
In particular, this information is required in 
order to relate the recent observation of $\tau$ 
production by the DONUT collaboration~\cite{donut} 
to the presence of \nutau in the beam.  
At 99\% C.L. the NOMAD data limit such couplings to  
$P_{\mu \tau} < 4.4 \times 10^{-4}$ and 
$P_{e\tau} < 2.0 \times 10^{-2}$.

\section{Conclusions} 
\label{sec:concl}

The analysis of the full NOMAD data sample  
gives no evidence for $\nutau$ appearance. 
In the two-family oscillation formalism this result 
excludes a region of the $\numunutau$ oscillation parameters which 
limits $\sin^2 2\theta_{\mu \tau}$ at high $\Delta m^2$ to values 
smaller than $3.3 \times 10^{-4}$ at 90\% C.L., and $\Delta m^2$ 
to values smaller than $\Delta m^2 < 0.7$ eV$^2$/$c^4$ at 
$\sin^2 2\theta_{\mu \tau}=1$.\ The corresponding excluded region 
at 90\% C.L. for the $\nuenutau$ oscillation parameters includes 
$\sin^2 2\theta_{e \tau} < 1.5 \times 10^{-2}$ at   
large $\Delta m^2$ and $\Delta m^2 < 5.9$ eV$^2$/$c^4$ at $\sin^2 
2\theta_{e \tau}=1$. Our sensitivity to oscillations 
is not limited by backgrounds, but is essentially defined 
by the available statistics.  

The NOMAD experiment has explored neutrino oscillations 
down to probabilities which are more than one order of 
magnitude smaller than limits set by the previous generation 
of experiments. For the first time, a purely kinematic approach 
has been applied to the detection of \nutau CC interactions.  
Our final results demonstrate that this approach has  
developed into a mature technique, providing a precise 
control of backgrounds from the data themselves.  \

\begin{ack}
We thank the management and staff of CERN and of all
participating institutes for their vigorous support of the experiment.
Particular thanks are due to the CERN accelerator and beam-line staff
for the magnificent performance of the neutrino beam. The following
funding agencies have contributed to this experiment:
Australian Research Council (ARC) and Department of Industry, Science, and
Resources (DISR), Australia;
Institut National de Physique Nucl\'eaire et Physique des Particules (IN2P3), 
Commissariat \`a l'Energie Atomique (CEA), Minist\`ere de l'Education 
Nationale, de l'Enseignement Sup\'erieur et de la Recherche, France;
Bundesministerium f\"ur Bildung und Forschung (BMBF, contract 05 6DO52), 
Germany; 
Istituto Nazionale di Fisica Nucleare (INFN), Italy;
Russian Foundation for Basic Research, 
Institute for Nuclear Research of the Russian Academy of Sciences, Russia; 
Fonds National Suisse de la Recherche Scientifique, Switzerland;
Department of Energy, National Science Foundation (grant PHY-9526278), 
the Sloan and the Cottrell Foundations, USA.  

We also thank our secretarial staff, Jane Barney, Katherine Cross, 
Joanne Hebb, Marie-Anne Huber, Jennifer Morton, 
Rachel Phillips and Mabel Richtering, 
and the following people who have worked with the 
collaboration on the preparation and the data 
collection stages of NOMAD:
M.~Anfreville, M.~Authier, G.~Barichello, A.~Beer, V.~Bonaiti, A.~Castera,
O.~Clou\'e, C.~D\'etraz, L.~Dumps, C.~Engster, 
G.~Gallay, W.~Huta, E.~Lessmann, 
J.~Mulon, J.P.~Pass\'e\-ri\-eux, P.~Petit\-pas, J.~Poin\-signon, 
C.~Sob\-czyn\-ski, S.~Sou\-li\'e, L.~Vi\-sen\-tin, P.~Wicht.
\end{ack}

\newcommand\bff{} 

\appendix
\section{Kinematic variables}
\label{sec:variables}

\begin{figure}[tb]
\begin{center}
  \epsfig{file=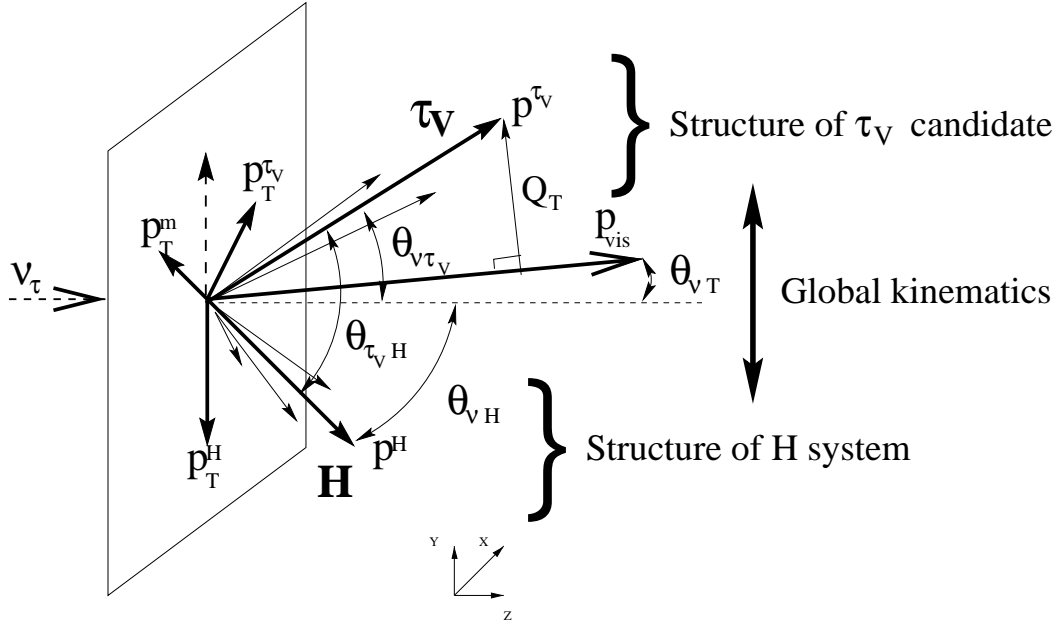,width=1.00\textwidth}
\end{center}
\caption{
Definition of the NOMAD kinematics for a $\nu_{\tau}$ CC event. 
}
\label{fig:kinevar}
\end{figure}

The event kinematics is based on a set of global variables 
which describe the general properties of the two momentum 
vectors of a {\em leading particle} and of the hadronic system $H$ 
recoiling against it, in the laboratory frame. This convention 
is closely related to the kinematics of CC interactions. 
However, in general, any track (or system of tracks) 
can be chosen as leading particle.  

Invariance with respect to an arbitrary rotation in the
plane transverse to the beam direction means that an
event can in fact be fully described by five degrees of
freedom (Figure~\ref{fig:kinevar}):
three in the transverse plane $(x,y)$
and two along the beam direction $(z)$.

In the selection of \nutau CC interactions, 
the leading particle consists of the vi\-si\-ble $\tau$ 
decay product(s) \xvis. The following kinematic variables 
can then be computed (Figure~\ref{fig:kinevar}):  
\begin{itemize}
\item $\mathbf p_{\rm vis}$, the total visible momentum of the event. 
      This is computed by summing the momenta of all primary 
      charged particles, neutral secondary vertices and 
      neutral ECAL clusters. 
\item $\evis$, the total visible energy of the event.

\item $\mathbf p^{\xvis}$ and $\mathbf p^{H}$, the total momentum of the 
{\em visible} tau decay product(s) and of the associated hadronic 
system respectively, such that $\mathbf p^{\xvis} 
+ \mathbf p^{H} = \mathbf p_{\rm vis}$. 

\item $\ybj$, the ratio between $p^{H}$ and the total visible energy.

\item $\mathbf{p}_T^{\,\xvis}$ and $\mathbf{p}_T^{\,\jet}$, the components 
of $\mathbf p^{\xvis}$ and $\mathbf p^{H}$ perpendicular to 
the neutrino beam direction. 

\item $\mathbf p_T^{\,m}$, defined as 
$-(\mathbf p_T^{\,\xvis}+\mathbf p_T^{\,\jet})$ 
and interpreted as a measurement of the ``missing'' transverse momentum 
due to the neutrino(s) from $\tau$ decay.

\item $M_T$, the transverse mass,  
given by $M_T^2 = 4p_T^{\xvis}p_T^m\sin^2(\phi_{\xvis m}/2)$ 
where $\phi_{\xvis m}$ is the angle between $\mathbf{p}_T^{\,\xvis}$
and $\mathbf p_T^{\,m}$, when assuming massless decay product(s). 
For $\tau$ events, $M_T \le$ $\tau$ mass, up to detector 
resolution and Fermi motion effects.

\item $R_{p_{T}}$, the ratio of the transverse momentum $p_T^{\xvis}$  
and the missing transverse momentum $p_T^m$, 

\item $Q_T$, the component of $\mathbf{p}^{\,\xvis}$ perpendicular to the total
visible momentum vector (including $\xvis$). 

\item $\qlep$, the component of a charged particle momentum perpendicular to
the total momentum of the rest of the event. 

\item $\theta_{\nu \jet}$, the angle between the neutrino beam direction
and the hadronic system.

\item $\theta_{\nu \xvis}$, the angle between the neutrino beam direction 
and the $\xvis$ momentum vector.

\item $\theta_{\nu T}$, the angle between the neutrino beam direction and the
total visible momentum vector of the event.

\item $\theta_{\xvis\jet}$, the angle between the hadronic system and $\xvis$.
\end{itemize}
For the lepton tagging algorithms (Section~\ref{sec:veto}) and 
the kinematic rejection of CC interactions (Section~\ref{sec:kine}) 
the lepton candidate track \lmucc is chosen as leading particle 
in the computation of all the above variables (e.g. $\theta_{\nu \xvis}$ 
becomes $\theta_{\nu l}$ etc.). 
 
The following variables, partially related to the previous ones, 
incorporate information from the internal structure of 
the hadronic system $H$: 
\begin{itemize} 
\item $\theta_{\xvis h_{i}}$, the minimum angle between $\xvis$  
and any other primary track $h_{i}$ in the event. 

\item $\langle Q_{T}^{2} \rangle_{\jet}$, average $Q_{T}^{2}$ computed among 
all the charged tracks of the remaining hadronic system $H$, after 
the exclusion of a particular track. This variable measures 
the transverse size of the hadronic system. 

\item $\langle Q_{T}^{2} \rangle_{T}$, average $Q_{T}^{2}$ computed among 
all the charged tracks in the event.  

\item $R_{Q_{T}}$, ratio between the transverse size of the 
hadronic system, $\langle Q_{T}^{2} \rangle_{\jet}$, and that 
of the full event, $\langle Q_{T}^{2} \rangle_{T}$. 
This variable is sensitive to the isolation of the 
particle(s) not included in the hadronic system $H$.  

\item $\Delta r_{\xvis h_{i}}= \sqrt{(\Delta 
\eta_{\xvis h_{i}})^{2}+(\Delta \phi_{\xvis h_{i}})^{2}}$, 
the minimum invariant opening cone between $\xvis$ and any other 
primary track $h_{i}$ in the event. This combines the differences 
of the corresponding angles in the transverse plane $\phi$ and 
of the pseudo-rapidity $\eta = -\ln \tan (\theta/2)$.  
\end{itemize} 
In addition, variables describing the internal structure of the 
candidate \xvis are used, where applicable, in order to increase 
background rejection:
\begin{itemize} 
\item $M_{\rho}$, invariant mass of a $\pi^{-}\pi^{0}$ combination. 

\item $M_{\rho^{0}}$, invariant mass of a $\pi^{-}\pi^{+}$ combination. 

\item $M_{\pi^{0}}$, invariant mass of a $\gamma\gamma$ combination. 

\item $M_{a_{1}}$, invariant mass of a $\pi^{-}\pi^{+}\pi^{-}$ combination. 

\item $\theta_{\pi^{-}\pi^{0}}$, opening angle between 
a $\pi^{-}$ and a $\pi^{0}$. 
The $\pi^{0}$ momentum can be obtained, in turn, from a single ECAL 
cluster ($1\gamma$) or from the sum of two separate ECAL clusters ($2\gamma$). 

\item $\theta_{\gamma\gamma}$, opening angle between two $\gamma$'s.  

\item $\theta_{\pi^{+}\pi^{-}}$, opening angle between a $\pi^{+}$ and 
a $\pi^{-}$. 

\item $\theta_{\pi^{-}\pi^{-}}$, opening angle between two 
distinct $\pi^{-}$'s. 

\item $E_{\pi^{0}}$, energy of a $\pi^{0}$ obtained from a single ECAL 
cluster ($1\gamma$) or from the sum of two separate ECAL clusters ($2\gamma$). 

\item $E_{\gamma}^{\rm max}$, maximum energy between two different 
$\gamma$'s used to reconstruct a $\pi^{0}$. 

\item $E_{\pi^{+}}$, energy of a $\pi^{+}$. 

\item $E_{\rho^{0}}$, energy of a $\pi^{+}\pi^{-}$ combination. 
\end{itemize} 
where the track charges refer to the $\tau^{-}$ selection  
(opposite for the $\tau^{+}$ selection).

\section{Additional topologies}
\label{sec:others}

In addition to the analysis of the hadronic DIS channels 
described in this paper, the \nutau 
appearance search in NOMAD includes the analysis of the  
\taue DIS decays and of the LM topologies from Ref.~\cite{PLB1}.  
The corresponding values of \ntaumu and \ntaue have been 
updated ac\-cor\-ding to the most recent beam 
predictions~\cite{nutaubeam}.  
Table~\ref{tab:others} summarizes all the relevant numbers 
for these samples, which supersede the ones 
quoted in Ref.~\cite{PLB1}.  

In principle, NOMAD is sensitive to all leptonic and hadronic
$\tau$ decay channels. However, the \taumu decay channel is
dominated by the \numu CC background where the primary muon
is positively {\em identified} in the detector. 
In view of the intrinsic difficulty of the evaluation 
of the systematic uncertainty on this background, due to 
the absence of a suitable control sample, this 
channel is not directly used for the \nutau search.

\begin{table}[tb]
\caption{
Number of background and data events in the signal region 
for the \taue DIS and LM topologies~\cite{PLB1}. 
The corresponding $\ntaumu$ and $\ntaue$, as defined in
Sections~\ref{sec:systau} and~\ref{sec:limits}, 
are listed in the last two columns.
The bins denoted by a star are considered as low background
bins (Section~\ref{sec:allcha}).
}
\label{tab:others}
\vspace*{0.2cm}
\begin{center}
\begin{tabular}{lccccccc}
  \hline
\multicolumn{2}{c}{Analysis} & Bin \# & Tot bkgnd & Data & $\ntaumu$ & $\ntaue$  & \\ \hline
$\taue$ & DIS & I   & $0.85\,^{+\,0.26}_{-\,0.16}$  &  2 & 143 & 2.8 & \\
\multicolumn{2}{c}{($\evis>12$ GeV)} & II   & $0.46\,^{+\,0.23}_{-\,0.12}$  &  1 & 136 & 2.9 & \\
        &     & III  & $0.18\,^{+\,0.18}_{-\,0.08}$  &  0 & 680 & 15.0 & * \\
        &     & IV   & $1.85 \pm 0.22$  &  2 & 554 & 14.0 & \\
        &     & V    & $0.78 \pm 0.15$  &  0 & 406 & 9.1 &  \\ 
        &     & VI   & $0.16 \pm 0.08$  &  0 & 1481 & 32.7 & * \\  
\multicolumn{2}{c}{($\evis<12$ GeV)} & I+IV+V & $0.77 \pm 0.26$ &  0 & 253 & 2.8 & \\ 
        &     & II+III+VI  & $0.27 \pm 0.13$  &  0 & 665 & 8.7 & * \\ \hline  
$\taue$ & LM & I   & $3.09 \pm 0.67$  &  3 & 282 & 2.9 &  \\
        &    & II   & $1.50 \pm 0.41$  &  2 & 286 & 2.9 & \\
        &    & III  & $0.82 \pm 0.41$  &  1 & 296 & 3.0 & \\ \hline 
$\tau \to h(n\pi^0)$ & LM & $\rho$ & $5.2 \pm 1.8$ &  7 & 480 & 8.9 & \\ 
  & LM & $h$  & $6.7 \pm 2.3$ &  5 &  377 &  7.8 & \\  
 $\tau \to 3h(n\pi^0)$ & LM & --  & $3.5\pm1.2$ & 5 & 298 & 5.2 & \\ \hline 
\end{tabular}
\vspace*{0.8cm}
\end{center}
\end{table}

\end{document}